\documentclass[11pt]{article}

\usepackage{jheppub}
\usepackage{multirow}
\usepackage{graphicx,psfrag, subfigure}
\usepackage{amsmath}
\usepackage{dsfont,caption}
\usepackage{setspace}
\usepackage{soul}
\numberwithin{equation}{section}
\usepackage{relsize}
\usepackage{hyperref}
\usepackage{amssymb}

\def\be{\begin{equation}}
\def\ee{\end{equation}}
\def\bea{\begin{eqnarray}}
\def\eea{\end{eqnarray}}

\begin{document}

\begin{titlepage}

\setcounter{page}{1} \baselineskip=15.5pt \thispagestyle{empty}

\bigskip\
\begin{center}
{\Large \bf A Simple System For Coleman-De Luccia}\\ \vspace{0.3cm}
 {\Large \bf Transitions}
\vskip 5pt
\vskip 15pt
\end{center}
\vspace{0.5cm}
\begin{center}
{Kate Eckerle$^{*,\dagger}$}

\end{center}\vspace{0.05cm}

\begin{center}
\vskip 4pt
\textsl{$^*$Dipartimento di Fisica, Universit\`a Degli Studi di Milano-Bicocca, Milan, Italy} \\ \vspace{0.2cm}
\textsl{$^\dagger$INFN, sezione di Milano-Bicocca, Milan, Italy}\\ \vspace{0.2cm}

\end{center}

\vspace{-0.5cm}
{\small \noindent \\
\begin{center}
	\textbf{Abstract}

\end{center}

This paper presents a simple framework that organizes thin-wall Coleman-De Luccia instantons based on the Euclidean geometries of their original and tunneled vacuum patches. We consider all a priori allowed vacuum pairs (de Sitter or Anti-de Sitter for either patch, Minkowski can be obtained as a limit of either), and $O(4)$-symmetric thin-wall geometries connecting them. For each candidate bounce geometry, either a condition under which a solution to the $O(4)$-invariant equations of motion exists is derived, or the would-be vacuum transition is ruled out. For the parameter regimes in which a solution exists, we determine whether  expansion/contraction of the bounce supplies a negative mode in the second variation of the Euclidean action. All results follow from the monotonicity of a single function.\noindent}

\vfill

\pagenumbering{gobble}
\begin{flushleft}
\vspace{0.8cm} \small \today
\end{flushleft}
\end{titlepage}
\tableofcontents

\newpage
\pagenumbering{arabic}

\section{Introduction}\label{intro}
\begin{figure}
 \centering
\includegraphics[width=.5\linewidth]{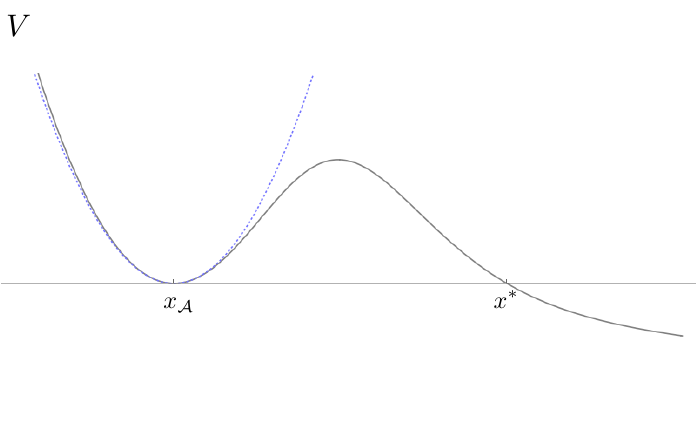}
\caption{\small The potential of a quantum mechanics system with a metastable state.}
 \label{fig-qm-V1}
\end{figure}

The hierarchy of gravitationally collapsed structures we observe in the universe could not have formed unless our dark energy density was small. Inflation isolated pockets of under and over density from one another, but had the cosmological constant (CC) been too large, the universe would have expanded too rapidly for those density perturbations  to seed the development of complex structures. This may help to explain why our observed vacuum energy density, $\rho_{\text{DE}}\sim 10^{-120}\thinspace\text{M}^4_{\text{pl}}$, is so far out of line with that of an ordinary effective field theory whose cut-off is at, or around the Planck scale. Effective field theory techniques predict vacuum energies that scale like $\hbar k^4_{\text{cut-off}}$, so $\rho_{\text{EFT}}\sim \mathcal{O}(1)\text{M}^4_{\text{pl}}$.  

Realizing an anthropic solution to the CC problem --- the idea that structure formation  itself, a seemingly  necessary condition for the act of observation, drastically limits the range of relevant CC values --- requires the effective field theory with Planck scale cut-off to have an enormous number of distinct vacuum phases \cite{Weinberg:1987dv,Bousso:2000xa}. Flux compactification of string/M-theory indeed suggests that quantum gravity may manifest in 4 dimensions at low energy as just such a landscape \cite{Giddings:2001yu, Strominger:1986uh, Denef:2004ze, Denef:2004cf, Denef:2004dm, Douglas:2006es, DDFGK}.  Whether the landscape includes \emph{tiny} positive values of the CC, like ours, depends on the distribution of vacuum energy densities across the landscape\footnote{We should remark, this string landscape picture does have skeptics. There are certain fundamental obstacles that stand in the way of identifying/explicitly constructing string vacua with specifically \emph{positive} CC \cite{Dine:1985he, Maldacena:2000mw}. This has lead some to  conjecture de Sitter (dS) vacua simply don't exist in the landscape \cite{Obied:2018sgi}. That said, confidence in explicit dS constructions, mainly  \cite{KKLT} which is based on lifting Anti-de Sitter vacua in a systematic way, has steadily risen. Recent discoveries of dS solutions in (classical)  massive IIA  supergravity involving orientifold planes also suggest the presence of dS in the landscape \cite{Cordova:2018dbb,Cordova:2019cvf}. Finally, the dS swampland conjecture itself was found to result in fine-tuning problems by about 55 orders of magnitude by \cite{Denef:2018etk}.}. 

A rough idea can be gotten by assuming a uniform distribution. The spacing between vacuum energies then is the inverse of the total number of vacua in the landscape, $1/\mathcal{N}_{\text{vac}}$, and so requisite number is $\mathcal{N}_{\text{vac}}\gtrsim 10^{120}$. Estimates of $\mathcal{N}_{\text{vac}}$ from string compactifications can easily exceed the ballpark value of $ 10^{120}$, essentially because of the fact that the relevant special geometry compact manifolds have $\gtrsim 100$ cycles, and the quanta of higher-form flux wrapping said cycles it seems can be varied independently, at least within certain regimes; say, over $10$ integers.

Vacua are configurations of the effective fields that locally minimize the potential (in the string/M-theory  context, vacua correspond to choices for the internal manifold's geometry, and configurations of the supergravity fields that together locally minimize the energy density). Though such configurations are perturbatively stable backgrounds to expand the effective fields about, they need not be  arbitrarily long-lived. Quantum fluctuations can cause a vacuum configuration to decay, loosely speaking via the fields within regions of spacetime --- ``bubbles" --- tunneling into other, new vacuum configurations. Vacua that are not absolutely stable are referred to as metastable. 

In an anthropic solution, many (if not all) vacuum phases are populated by metastable decay, and/or consequences thereof; for instance, the collision of multiple bubbles nucleated at different spacetime locations \cite{Easther_2009,Giblin_2010,Ahlqvist_2015}. Most spacetime patches contain no observers because $\rho_{\text{DE}}$ is outside the anthropically acceptable range. At the same time, it is not necessarily the case that the remaining regions of small CC \emph{do} contain observers.

The reason is because phase transitions to low CC vacua may or may not result in initial conditions consistent with structure formation. For example,  in a generic landscape bubble nucleation is not followed by a period in which the spacetime in the interior of the bubble inflates by a  sufficient amount  to reduce the spatial curvature inside the bubble to the degree required for structure formation to follow \cite{Freivogel:2005vv}. Initial conditions can be influenced directly by metastable vacuum decay processes, like in this example (bubble nucleation), or indirectly via the consequences of successive decay events (like bubble collisions \cite{Kleban:2011pg, Gobbetti:2012yq, Aguirre_2011}).

In addition to understanding the decay channels \emph{to}  low CC vacua, we also need to understand those \emph{from} low CC vacua. This is necessary because in addition to ensuring enough inflation, we separately need to ensure the low CC universe survives for, in our case, 13.8 billion years. These examples illustrate that the viability of an anthropic solution is a comprehensive assessment that takes into account both the landscape under consideration, and the mechanism(s) that populate its different phases. \emph{How} phase transitions unfold  in a particular landscape, whether generated by string compactification or otherwise, determines whether conditioning on structure formation renders our CC (and a universe like our own more broadly) natural. Metastable vacuum decay particularly within the context of dynamical spacetime is an essential piece of the puzzle, and is the subject we examine in this paper.

Metastable states can be studied using the Euclidean path integral. These methods enable the calculation of a semi-classical  decay rate $\Gamma$ of the form,
\be
\Gamma = Ae^{-B/\hbar}(1+\mathcal{O}(\hbar)).\label{eq-Gamma-intro}
\ee
The coefficients, $A$, and $B$, are defined in terms of a solution to the Euclidean equations of motion known as ``the bounce"  \cite{Coleman:1977py, Callan:1977pt}. The bounce is a non-constant solution with maximal symmetry that approaches the false vacuum asymptotically, and explores only the barrier region. In the cases of particle tunneling in quantum mechanics (including theories of multiple interacting particles), and vacuum decay in quantum field theory in flat space (for theories involving scalars and/or gauge fields, and given certain assumptions on the potential) the bounce is known to be a degree-1 saddle point of the Euclidean action, $S_{\text{E}}$  \cite{ Callan:1977pt,Coleman:1987rm, Coleman:1978ae}. In other words, the differential operator appearing in the second variation of the Euclidean action evaluated at the bounce has exactly one negative mode.

In a nutshell, Euclidean paths located a small distance away from the bounce in this negative mode direction produce an imaginary contribution to the false-vacuum-to-false-vacuum Euclidean propagator, $K_{\text{E}}$, when the technique of steepest descent is applied to the path integral expression for $K^{}_{\text{E}}$. This imaginary term exponentiates once the analogous leading order contributions to $K_{\text{E}}$ from the vicinities of multi-bounce paths are included. The decay rate is associated specifically to the contributions from this subspace of the path integral's domain; namely to the asymptotics of their sum's logarithm.

The result is an expression of the form \eqref{eq-Gamma-intro}, where $B$ is the difference in Euclidean action between the bounce and static false vacuum path, and $A$ is a dimensionful factor involving the spectra of the second variational derivative of $S_{\text{E}}$, evaluated at the bounce, vs evaluated at the static false vacuum path. This calculation is reviewed in detail in appendix \ref{appendix-qm} for the case of particle tunneling in quantum mechanics, with particular attention to the use of steepest descent.

For a system that is initially in an approximately stationary state, the bounce computes a rate reflective of how the time-evolved probability amplitude  leaks through a finite height potential barrier. For example, for the quantum particle system shown  in figure \ref{fig-qm-V1}, the bounce tells us about the dynamics of the ground state of the related harmonic oscillator system when it is evolved by the actual system's Hamiltonian. The bounce is accurate in describing the loss in probability from the well region, $\int_{x^\ast}^\infty dx |\psi(t,x)|^2\approx 1- e^{-\Gamma t}$, during the initial time period, before any non-decay phenomena  like reflection of the probability amplitude back into the well  have time to occur.

Conservation of $V-\frac{m}{2}\dot{x}^2$ in the Euclidean theory, where the dot refers to derivatives with respect to Euclidean time, $\tau=-it$, implies that $B$ works out to 
\be
B=\int_{x_{\mathcal{A}}}^{x^\ast} dx \sqrt{2m(V(x)-V_{\mathcal{A}})},
\ee where $x_{\mathcal{A}}$ is the position of the local minimum, $V_{\mathcal{A}}$ is its potential, and $x^\ast$ is the point of equipotential with $x_{\mathcal{A}}$ across the barrier. The fact that the path integral calculation we've described recovers the correct exponential dependence of the transmission coefficient computed in the WKB approximation is strong evidence for trusting instanton methods, at least within certain parameter regimes, to characterize decay phenomena. We briefly review the extension of these methods to field theory (which some readers may wish to skip) before presenting a simple means of organizing Coleman-De Luccia (CDL) decays, the purpose of this paper.

\subsection{Field theory}\label{subsec-flat}

The formulation of the particle's WKB/semiclassical tunneling rate in terms of a path integral is powerful because it can be generalized to field theory, including the case where the spacetime metric is allowed to fluctuate along with matter fields \cite{Coleman:1980aw}. Vacuum decay in field theory is  still fundamentally a tunneling phenomenon. In quantum mechanics, fluctuations through a potential barrier in position space mediate transitions out of a false vacuum state, while in field theory it is fluctuations in the field configuration(s) --- nevertheless through potential barriers --- that mediate transitions. The association of an imaginary contribution to $K_{\text{E}}$ from the vicinity of the analogous degree-1 saddle points of $S_{\text{E}}$ in field theory, instantons, carries over. This subsection reviews how things work in the context of field theory in flat space. In subsection \ref{subsec-intro-grav} we review the approach of CDL to incorporating gravitational effects, and set-up the framework used throughout the rest of this paper. 

An instanton is a non-constant solution of the Euclidean equations of motion that starts and ends in the false vacuum. The Euclidean path integral computation in field theory assigns a semi-classical decay rate to the metastable vacuum configuration of identical form as for the quantum particle, 
\be
\Gamma=Ae^{-B/\hbar} \left(1+\mathcal{O}(\hbar)\right).
\ee
$B$ is still the difference in Euclidean actions between the instanton and static false vacuum state,
\be
B=S_{\text{E, inst}}-S_{\text{E, f.v.}}
\ee
where $S_{\text{E}}$ is now a functional of field configuration(s), while the field theory generalization of $A$ is,
\be
A= \frac{B^2}{4\pi^2\hbar^2} \left(\sqrt{\frac{\text{det}(S''_{\text{E, f.v.}})}{\text{det}'(S''_{\text{E, inst}})}}\right)\label{eq-A-coeff},
\ee
where $\det'$ signifies the product of the magnitudes of nontrivial eigenvalues. $A$ has dimensions of inverse 4-volume because of the square-root factor.  The bounce in flat space has four zero modes, corresponding to shifts of its center in $\mathbb{R}^4$. Therefore, $\text{det}'(S''_{\text{E, inst}})$ contains four fewer eigenvalues relative to $\text{det}(S''_{\text{E, f.v.}})$. The eigenvalues of the operator appearing in $S''_{\text{E}}$, for instance in a single scalar field theory expanded about background $\phi$ the operator,
\be
- \Delta_{(4)} +V''(\phi),\label{eq-operator}
\ee
have units of inverse length squared, so the square-root factor in \eqref{eq-A-coeff} indeed has units of $\text{length}^{-4}$.  $\Delta_{(4)}$ in \eqref{eq-operator} denotes the Laplacian in $\mathbb{R}^4$. 

These identifications for $A$ and $B$ hold in so far as a saddle approximation to the path integral is valid, and the instanton has one negative mode. How to handle the case of multiple negative modes, in general, is unclear. Some believe these never describe decay channels because they attribute  one factor of $i$ to each negative mode of the single-bounce contribution to $K_{\text{E}}$. 
It seems at least conceivable that the application of steepest descent for multi-negative mode single-bounces is more subtle, and that correctly doing so resolves the matter by producing an overall phase factor of $i$ instead of $i^k$, where $k$ is the number of negative modes per single-bounce. In fact, if it is truly the case that the $i^k$ treatment of degree-$k$ single-bounce saddles is correct, $k>1$ that happen to have $k \pmod 4=1$ would be perfectly acceptable.  This condition strikes as somewhat arbitrary. Yet, it's likewise possible that this skepticism is misguided, so we take an agnostic position on whether multi-negative mode single-bounces describe decay. A simple way for all degree  $k>1$ instantons in a theory to not be associated with decay channels would be for the steepest descent contour to not encounter them. 

What \emph{is} typically viewed as a safe assumption is that the lowest Euclidean action instanton describes the dominant phase transition out of a given vacuum, since a more subtle application of steepest descent would nevertheless produce a factor of $e^{-B/\hbar}$ relative to the zero point correction to the false vacuum energy. The lowest Euclidean action instanton is typically expected to be that with maximal symmetry. This is proven for a scalar field $\phi$ in Minkowski space (given certain assumptions on the potential), where the bounce is an $O(4)$-symmetric extremum of $S_{\text{E}}$ \cite{Coleman:1977th}. The $O(4)$-symmetric bounce indeed has one negative mode, which itself is $O(4)$-invariant. This radial mode corresponds to expanding/contracting the bounce, just as the negative mode does in the quantum mechanics case.

Concretely, the generalization of the bounce to a theory involving a single scalar $\phi$ in flat space,
\be
S=\int d^4 x \thickspace \frac{1}{2}\left(\frac{\partial\phi}{\partial t}\right)^2 - \frac{1}{2}|\nabla \phi|^2-V(\phi)
\ee
goes as follows. The equation of motion in $\tau=-it$ reads,
\be
\Delta_{(4)} \phi=\frac{d V}{d\phi}.
\ee
$O(4)$ symmetry of the bounce is imposed by taking $\phi(\tau, \vec{x})=\phi(\rho)$ where $\rho$ is the radial coordinate of the $\mathbb{R}^4$ in spherical coordinates $\rho=\sqrt{\tau^2+|\vec{x}|^2}$,
\be
ds_{\text{E}}^2=d\rho^2 + \rho^2 d\Omega^2\label{euc-metric-flat},
\ee
$d\Omega$ is the line element on $S^3$. The Euclidean equation of motion becomes,
\be
\phi''+\frac{3}{\rho}\phi'=\frac{d V}{d\phi},\label{euc-eom-flat}
\ee
with primes denoting derivatives with respect to $\rho$. 

Let the potential $V$ have a local minimum at $\phi_{\mathcal{A}}$. 
The bounce describing decay out of $\phi_{\mathcal{A}}$ is a solution to \eqref{euc-eom-flat} that satisfies the boundary conditions,
\begin{align}
\lim_{\rho\rightarrow \infty}\phi(\rho)&=\phi_{\mathcal{A}}\\
\phi'(0)&=0.
\end{align}
The center of the bounce is at $\rho=0$; it is the analog of $T/2$ in the quantum mechanics case discussed in appendix \ref{appendix-qm}. These boundary conditions only permit a non-trivial solution if the  velocity in the infinite past (whose magnitdue is vanishingly small) points in the direction of a finite height potential barrier.

Equation \eqref{euc-eom-flat} is the equation of a classical particle in one dimension rolling in the inverted potential, under the influence of a damping term. The position of this auxiliary particle is $\phi$, the time of the auxiliary theory is $\rho$, and the coefficient of the damping term is ``time"-dependent, $\frac{3}{\rho}$. (the damping term is simply an artifact of writing the 4-dimensional Laplacian  in spherical coordinates.)

The qualitative features of the bounce are similar to those in the quantum mechanics case. There is a unique location $\phi^\ast$ on the other side of the barrier such that, if released from this position at $\rho=0$ with zero velocity, the auxiliary particle rolls through the valley of $-V$   in finite time and comes to rest at $\phi_{\mathcal{A}}$ asymptotically. The $\phi^\ast$ is strictly beyond the point of equipotential with $\phi_{\mathcal{A}}$ due to the fact that the particle loses energy because of the damping term. The trajectory obtained by even reflection across $\rho=0$, i.e. a full roundtrip, describes a particle that comes in from $\phi_{\mathcal{A}}$ from the infinite past, bounces off the barrier at location $\phi^\ast$ at time $\rho=0$, and returns to $\phi_{\mathcal{A}}$ in the infinite future.

\begin{figure}
 \centering
\includegraphics[width=1\linewidth]{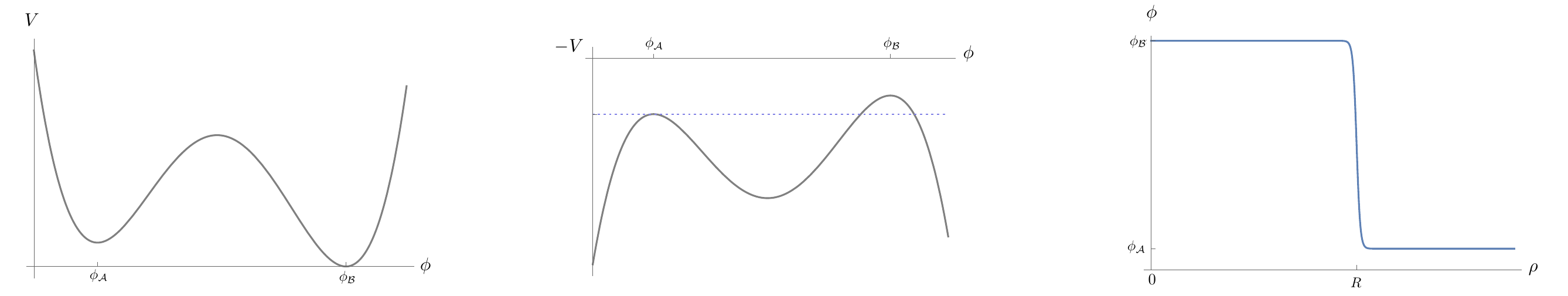}
\caption{\small Left: double-well potential. Middle: Inverted potential. Right: thin-wall bounce for the potential pictured. }
 \label{fig-thin-wall-panel}
\end{figure}

There is a particular parameter regime that can be studied analytically, namely that where a true vacuum  $\phi_{\mathcal{B}}$ lies on the other side of the barrier, and the barrier is high/broad relative to the difference in vacuum energies $\Delta V=V_{\mathcal{A}}-V_{\mathcal{B}}$. The damping term in \eqref{euc-eom-flat} turns out to have a small effect in this setting. Qualitatively, the bounce trajectory consists of two exponentially flat portions --- one of finite length $R$ with $\phi$ approximately in $\phi_{\mathcal{B}}$, and the other of arbitrarily large size with $\phi$ approximately equal to $\phi_{\mathcal{A}}$ --- joined by a smooth, monotonic function of width $w\ll R$. The wall profile is well-approximated by the soliton of an associated theory in which the vacua are exactly degenerate (the soliton should be scaled and shifted to interpolate between the $\phi$ values of the non-degenerate theory's local minima). An example is shown in figure \ref{fig-thin-wall-panel}.   

For such a  piece-wise defined $\phi$, $B$ can be estimated as,
\begin{align}
B&=2\pi^2 \int_0^R d\rho \thinspace \rho^3 \left(\frac{1}{2}\phi'^2+V(\phi)-V_{\mathcal{A}}\right)\\
&\approx \frac{\pi^2}{2}R^4 (V_{\mathcal{B}}-V_{\mathcal{A}})+ 2\pi^2 \sigma  R^3 .\label{B-tw-flat}
\end{align}
where $\sigma$ is the action of the aforementioned 1-dimensional soliton,
\be
\sigma=\int_{\phi_1}^{\phi_2}d\phi \sqrt{2(U(\phi)-U_0)}.
\ee
$U$ is the associated degenerate theory's potential, and $\phi_1$ and $\phi_2$ are the locations of its two minima, $U(\phi_1)=U(\phi_2)=U_0$. Only $R$ values that extremize $B$ in \eqref{B-tw-flat} correspond to solutions of the Euclidean equation of motion. 
\be
B'=-2\pi^2 R^3 \Delta V + 6\pi^2 \sigma R^2 =0
\ee
 has one nontrivial solution, $R=3\sigma/\Delta V$, where $\Delta V=V_{\mathcal{A}}-V_{\mathcal{B}}$.  This extremum is a maximum of \eqref{B-tw-flat}, and is referred to as the thin-wall radius. Varying $R$, which grows or contracts the bounce, corresponds to going \emph{off}-shell, albeit in a manner that maintains $O(4)$ symmetry and interpolates back and forth between the two vacua. The fact that this deformation away from the bounce decreases the Euclidean action reflects the presence of a negative mode of $S_{\text{E}}''$ evaluated at the bounce. The spectrum, in fact, contains only one negative mode, and it is this $O(4)$-symmetric deformation associated with expansion/contraction of the bounce, the result we mentioned a moment ago. 
 
When the bounce is continued to Lorentzian time, it takes the form of a spatially spherically symmetric field configuration, a 3-dimensional bubble of $\phi\approx \phi_{\mathcal{B}}$ separated from a surrounding region in $\phi_{\mathcal{A}}$. The thin-wall radius $R$ is the radius of the bubble at its moment of nucleation. The bubble wall nucleates at rest, and subsequently accelerates outward, converting an ever-growing surrounding region of $\phi_{\mathcal{A}}$ into $\phi_{\mathcal{B}}$. $O(4)$ symmetry translates to $SO(3,1)$ invariance of the wall's trajectory.

In general, the width of a soliton is given by integrating $\frac{1}{\sqrt{2(U(\phi)-U_0)}}$ over a $\phi$-interval between $\phi_1$ and $\phi_2$ that contains a significant enough fraction of the barrier's height given one's standard of accuracy; ``standard of accuracy" meaning how closely one requires $U(\phi_{\text{soliton}}(\pm w))$ to be to $U_0$   to be considered outside of the wall. For a barrier of height $H=U_{\text{peak}}-U_0$ that is well-described by a quadratic, the width can be estimated by, 
\begin{align}
w&\approx2\int_0^{a} d\phi \left(2(H-\frac{1}{2}U''_{\text{peak}}\phi^2 )\right)^{-1/2},\quad \text{where }a=\sqrt{2H/U''_{\text{peak}}}\\
&=\frac{\pi}{2 \sqrt{U''_{\text{peak}}}}.
\end{align}
Using the quadratic approximation for $U$ in the soliton action, on the other hand, gives 
\be
\sigma\approx  \frac{H\pi}{2\sqrt{U''_{\text{peak}}}}.
\ee
Therefore, an estimate of $\frac{w}{\sigma}$ is the inverse of the barrier height, $\frac{1}{V_{\text{h}}}$, and the regime of validity of the thin-wall approximation is $V_{\text{h}}\gg\Delta V$.

\subsection{Bounces in gravity}\label{subsec-intro-grav}
To allow for gravitational effects, begin by adding to the scalar action the Einstein-Hilbert term and requisite boundary terms,
\be
S=\int d^4 x \sqrt{-g} \left(\frac{1}{2}\partial_\mu\phi\partial^\mu\phi-V(\phi)-\frac{1}{16\pi G}R\right)\label{action-lor-sig}+\text{BTs}
\ee
where $G$ is Newton's constant, and $R$ is the curvature scalar of the spacetime defined by metric $g_{\mu \nu}$ whose signature is $+---$. 
We make the same ansatz of $O(4)$ symmetry for the lowest action instanton. The most general Riemannian metric of this form is\footnote{Note that there is a gauge degree of freedom associated with rescalings of $\xi$.},
\be
ds_{\text{E}}^2=d\xi^2+\rho(\xi)^2d\Omega^2\label{euc-metric}.
\ee
To impose $O(4)$ symmetry take,
\be
\phi=\phi(\rho(\xi)).
\ee

Evaluating the Euclidean action on an $O(4)$-symmetric configuration, and integrating over the 3-sphere yields, 
\be
S_\text{E, eff}=2\pi^2\int d\xi \left(\rho^3\left(\frac{1}{2}\phi'^2+V\right)+\frac{3}{\kappa}\left(\rho^2\rho''+\rho \rho'^2-\rho\right)\right)\label{euc-action}
\ee
where we have excluded boundary terms, and used $\kappa\equiv8\pi G$. Only two of the Euclidean equations of motion are independent once $O(4)$ symmetry is imposed. They can be expressed as, 
\begin{align}
&\rho'\thinspace^2=1+\frac{1}{3}\kappa\rho^2\left(\frac{1}{2}\phi'^2-V(\phi)\right)\label{phi-EOM}\\
&\phi'' +\frac{3\rho'}{\rho}\phi'=\frac{d V}{d\phi}\label{rho-EOM}
\end{align}
where primes indicate derivatives with respect to $\xi$.
We are only ever interested in $B= S_{\text{E, inst}}-S_{\text{E, } \mathcal{A}}$, hence any boundary terms from \eqref{euc-action} cancel. The action \eqref{euc-action} evaluated on a solution to \eqref{phi-EOM} and \eqref{rho-EOM} (excluding boundary terms) is given by,  
\be\label{SE-on-shell}
S_{\text{E, on-shell}}=4\pi^2\int d\xi \left(\rho^3 V-\frac{3\rho}{\kappa}\right).
\ee 

Vacuum solutions are those with a homogeneous $\phi$ that locally minimizes $V$, and a $\rho$ that solves,
\be
\rho'^{\thinspace2}=1-\frac{\kappa V_{\text{vac}}}{3}\rho^2.\label{eq-vacuum-rho}
\ee
Defining $\ell^2=(3/\kappa |V_{\text{vac}}|)$, \eqref{eq-vacuum-rho} can be expressed as,
\be
 \rho'^{\thinspace2}\pm \rho^2/ \ell^2 = 1\label{eq-vacuum-rho-2},
\ee
where the sign is given by whether the vacuum energy is positive or negative --- plus for $V_{\text{vac}}>0$, minus for $V_{\text{vac}}<0$. For the first case, de Sitter (dS),  \eqref{eq-vacuum-rho-2} is the equation of the unit circle. Setting the origin of the Euclidean signature spacetime at $\xi=0$, we find the solution $$\rho_{\text{dS}}(\xi)=\ell \sin(\xi/\ell).$$ The full metric \eqref{euc-metric} is therefore that of a $4$-sphere with radius $\ell$. 

Keeping the choice $\xi=0$ for the origin we find the solution  $$\rho_{\text{AdS}}(\xi)=\ell\sinh(\xi/\ell)$$ for negative vacuum energies, and therefore recognize Euclidean Anti-de Sitter (AdS) as a hyperboloid. Whereas the range of $\xi$ for dS is $[0,\pi\ell]$, that for AdS is $\mathbb{R}_+$. For the Minkowski case, the solution to \eqref{eq-vacuum-rho} is $\rho(\xi)=\xi$, and we recover the spherically symmetric flat metric \eqref{euc-metric-flat} (i.e. $\rho=\xi$ is the radius in spherical coordinates on $\mathbb{R}^4$). In our paper's analysis, it suffices to consider only dS and AdS vacua for parent/daughter vacua because Minkowski vacua can be obtained as a limit of either; $V_{\text{dS}}\rightarrow 0^+$, or $V_{\text{AdS}}\rightarrow 0^-$.

Thin-wall field configurations consist of pairs $(\phi,\rho)$ with $\phi$ resembling the approximately piece-wise constant (but continuous and monotonic) functions we described in subsection \ref{subsec-flat},
\begin{align}
&\nonumber\\
\phi(\rho;R)\approx\begin{array}{cc}
  \Bigg\{& 
    \begin{array}{cc}
     \phi_{\mathcal{B}} &\thickspace 0<\rho\lessapprox R \\
    \phi_{\text{w}}(\rho)   &\qquad\rho\approx R  \\
      \phi_{\mathcal{A}} & \qquad\rho\gtrapprox R
    \end{array}\label{eq-phi-pcwise}
\end{array}
\end{align}
where $ \phi_{\text{w}}$ is a soliton-like function centered at $R$, like the one pictured in figure \ref{fig-thin-wall-panel}, along with an approximately piece-wise defined $\rho(\xi)$, 
\begin{align}
&\nonumber\\
\rho(\xi)\approx\begin{array}{cc}
  \Bigg\{& 
    \begin{array}{cc}
     \rho_{\text{vac}}(\xi;V_{\mathcal{B}}) &\thickspace 0<\xi\leq \xi^\ast \\
       \rho_{\text{vac}}(\xi-\delta;V_{\mathcal{A}})& \qquad\xi\geq \xi^\ast
    \end{array}
\end{array}\label{eq-rho-pcwise}
\end{align}
where $\rho(\xi^\ast)=R$. On either side of the wall, the radial metric function $\rho(\xi)$ takes the vacuum solutions because $\phi'\approx 0$ and $V\approx V_{\mathcal{A}} $ or $V_{\mathcal{B}}$. The spacetime manifold in Euclidean signature needs to be connected, so a shift in the argument $\xi$ in \eqref{eq-rho-pcwise} is required so that $\rho$ is continuous. $R$ is viewed as a parameter that fixes $\xi^\ast$ via $ \rho_{\text{vac}}(\xi^\ast;V_{\mathcal{B}})=R$. Note that for the $V_{\mathcal{B}}>0$ case, there are two choices for $\xi^\ast$ allowed, a priori. $\xi^\ast$ can lie in the northern hemisphere, $0<\xi^\ast/\ell_{\mathcal{B}}<\pi/2$, or the southern, $\pi/2<\xi^\ast/\ell_{\mathcal{B}}<\pi$. The coordinate shift $\delta$ is fixed by $\rho_{\text{vac}}(\xi^\ast-\delta;V_{\mathcal{A}})=R$. Likewise, $\delta$  has two choices if $V_{\mathcal{A}}$ is positive. For example,  one option for a dS-to-dS configuration consists of a less-than-quarter-wavelength  oscillation of the $V_{\mathcal{B}}$ sine function starting at the origin, connected to a less-than-quarter-wavelength oscillation of the $V_{\mathcal{A}}$ sine function. The different options for $\xi^{\ast}$ and $\delta$ correspond to whether less, or more, than a hemisphere of the $\mathcal{B}$ dS is connected to less, or more, of a hemisphere of the $\mathcal{A}$ dS. 

A patched field configuration $(\phi,\rho)$ that corresponds to a solution to the Euclidean equations of motion may, or may not exist given the values of the  two vacuum energies,  $V_{\mathcal{A}} $ and $V_{\mathcal{B}}$, the wall tension $\sigma$, and the option/type of patched spacetime geometry implied by $\xi^\ast$ and $\delta$. We will determine which parameter regimes allow for an $O(4)$-symmetric bounce by determining when $B'(R)=0$ has a nontrivial solution,  for all choices of vacuum pairs and options for the instanton geometry.  

To that end, we define a function that captures the contribution to $S_{\text{E}}$ from vacuum patches. For an AdS vacuum,  $\rho'$ is positive for all $\xi$ so we take the positive square root of \eqref{eq-vacuum-rho} to relate $d\xi$ and $d\rho$ by,
\be
d \rho=\sqrt{1-\frac{\kappa V_{} \rho_{\text{AdS}}^2}{3}}\thinspace d\xi.
\ee
We therefore obtain the contribution
\begin{align}
F_1(V,R)&\equiv4\pi^2 \int_0^{\xi^*} d\xi \left(\rho^3_{\text{AdS}}(\xi) V_{\mathcal{A}}-\frac{3\rho_{\text{AdS}}(\xi)}{\kappa}\right)\\
&= -\frac{12\pi^2}{\kappa} \int_0^{R} d\rho   \thinspace \rho\sqrt{1-\rho^2 \frac{\kappa V}{3} }\\
&=\frac{12 \pi^2}{\kappa^2 V}\left( \left(1-R^{2} \frac{\kappa V}{3}\right)^{3/2}-1\right).\label{eq-cap-fn}
\end{align}
Since $V<0$ in AdS, this is equivalent to $F_1(V,R)=-\frac{12 \pi^2}{\kappa^2 |V|}\left( \left(1+\frac{R^{2}}{\ell_{\text{AdS}}^2}\right)^{3/2}-1\right)$.

There are two types of dS patches --- a less than a hemisphere-sized patch, or a more than a hemisphere-sized patch. We refer to the former as a ``cap" of the dS, and the latter as a ``bulb." The contribution to $S_{\text{E}}$ from a cap takes identical form to that of the AdS patch because $\rho_{\text{dS}}'$ is positive in the northern hemisphere. The only difference for a dS cap is that $F_1(V,R)$ is evaluated with positive $V$, resulting in $$F_1(V,R)=\frac{12 \pi^2}{\kappa^2 V}\left( \left(1-\frac{R^{2}}{\ell_{\text{dS}}^2}\right)^{3/2}-1\right).$$ We refer to $F_1$ in equation \eqref{eq-cap-fn} as the cap function regardless of the sign of $V$ with which it is evaluated. 

The contribution from a bulb of dS is simply the Euclidean action of the full 4-sphere minus the contribution from a cap,
\begin{align}
F_2(V,R)&=S_{\text{E}}[\text{full dS}]-F_1(V,R)\\
&=2 F_1(V,\ell_{\text{dS}})-F_1(V,R)\\
&=-\frac{24 \pi^2}{\kappa^2 V}-F_1(V,R).
\end{align}
This follows from breaking the integral \eqref{SE-on-shell} into a contribution from the northern hemisphere where $\xi/\ell\in [0,\pi/2]$ and $\rho'>0$, and the contribution from the remaining interval in the southern hemisphere $\xi/\ell\in [\pi/2,\xi^*/\ell]$, where $\rho'<0$\footnote{Trade integration variables using the short-cut provided by \eqref{rho-EOM} while properly taking into account the sign of $\rho'$:\begin{align*}
F_2(V_{\mathcal{A}},R)&=-\frac{12\pi^2}{\kappa}\int_0^{\ell_{\text{dS}}}d\rho \thinspace \rho\sqrt{1-\rho^2 \kappa V/3 }+\frac{12\pi^2}{\kappa}\int_{\ell_{\text{dS}}}^{R}d\rho \thinspace \rho\sqrt{1-\rho^2 \kappa V/3 }\\
&=F_1(V_{\mathcal{A}},\ell_{\text{dS}})+\frac{12\pi^2}{\kappa}\int_{\ell_{\text{dS}}}^{R}d\rho \thinspace \rho\sqrt{1-\rho^2 \kappa V/3 }\\
&=-\frac{12\pi^2}{\kappa^2 V}\left( 1+\left(1-\frac{\kappa V R^{2}}{3}\right)^{3/2}\right)
\end{align*}
}.

Finally, we define the square-root function appearing in $F_1(V,R)$ as follows,
\begin{align}
f(V,x)&= \sqrt{1-\kappa V x^2/3}\\
F_1(V,x)&= \frac{c_1}{V}\left( f(V,x)^3-1\right)
\end{align}
where $c_1$ is the constant $\frac{12\pi^2}{\kappa^2}$.  The purpose of this paper is to organize the different kinds of thin-wall CDL instantons within a  framework that simplifies their analysis. In order to consider uptunneling and downtunneling transitions systematically, we adopt the convention that the $\xi<\xi^\ast$ region of a thin-wall configuration always signifies the new --- or, ``daughter" --- vacuum configuration, while $\xi>\xi^\ast$ signifies the original ---or, ``parent" --- vacuum configuration. $\mathcal{A}$ subscripts label the parent, and $\mathcal{B}$ subscripts label the daughter, \emph{regardless} of which has higher energy density.

For all choices of parent-daughter vacuum pairs, and a priori allowed thin-wall bounce geometries, we first identify the subset of $4$-geometries that can ever correspond to an $O(4)$-symmetric extremum of $S_{\text{E}}$. Then we derive conditions on the relative values of the vacuum energies and thin-wall tension in order for the solution to exist. The conditions take the form of inequalities; for each type of bounce they separate the $(V_{\mathcal{A}}, V_{\mathcal{B}}, \sigma)$ parameter space into a region for which a nontrivial solution to $B'(R)=0$ exists, from the region for which it does not. These inequalities follow from the monotonicity properties of $f(V,x)$ (for positive $V$, and for negative $V$) alone. For dS-dS bounces we recover the distinction between type-A and type-B instantons, wherein the former have the traditional $O(4)$-invariant negative mode associated with expansion/contraction of the bounce, and the latter do not. 

Our goal is to apply this framework in the future to study aspects of CDL transitions whose understanding is less straightforward. These include the potential recovery of $O(4)$-invariant negative modes for type-B instantons, evidence of which was found can occur by \cite{Lee:2014uza} and \cite{Yang:2012cu}, and the possible resolution of obscurities arising in the decay rates of dS parents. 

There are ambiguities  interpreting the negative modes of gravitational instantons. Originally, \cite{Lavrelashvili:1985vn} found that gravity seems to allow for bounces with an infinite tower of negative modes, and argued that this indicates the breakdown of the WKB approximation. \cite{Tanaka:1992zw} used both a Hamiltonian formalism and the Wheeler-De Witt equation to argue that the breakdown of WKB was superficial, and the result of \cite{Lavrelashvili:1985vn}'s choice of gauge. It was further argued in \cite{Tanaka:1999pj} that this Hamiltonian formulation of the action implied there were no negative modes associated with gravitational degrees of freedom. The Hamiltonian formalism was used by \cite{Gratton:2000fj} to clarify the matter of negative modes for a class of  CDL, and Hawking-Moss instantons, finding a single negative mode was present. \cite{Gratton:2000fj} also recover a negative mode for Hawking-Turok constrained instantons by using a particular regularization. 

Many groups \cite{Lavrelashvili:1999sr, Khvedelidze:2000cp, Koehn:2015hga, Bramberger:2019mkv, Hertog:2018kbz, Gratton:2001gw, Battarra:2012vu} have studied issues involving the negative modes of gravitational instantons, and clarified different aspects of the matter. Nevertheless, how we ought to interpret gravitational instantons, especially in describing decays from dS parents, has not been definitively settled. 

\section{dS-dS transitions}\label{sec-dS-dS}

\begin{figure}
 \centering
\includegraphics[width=1\linewidth]{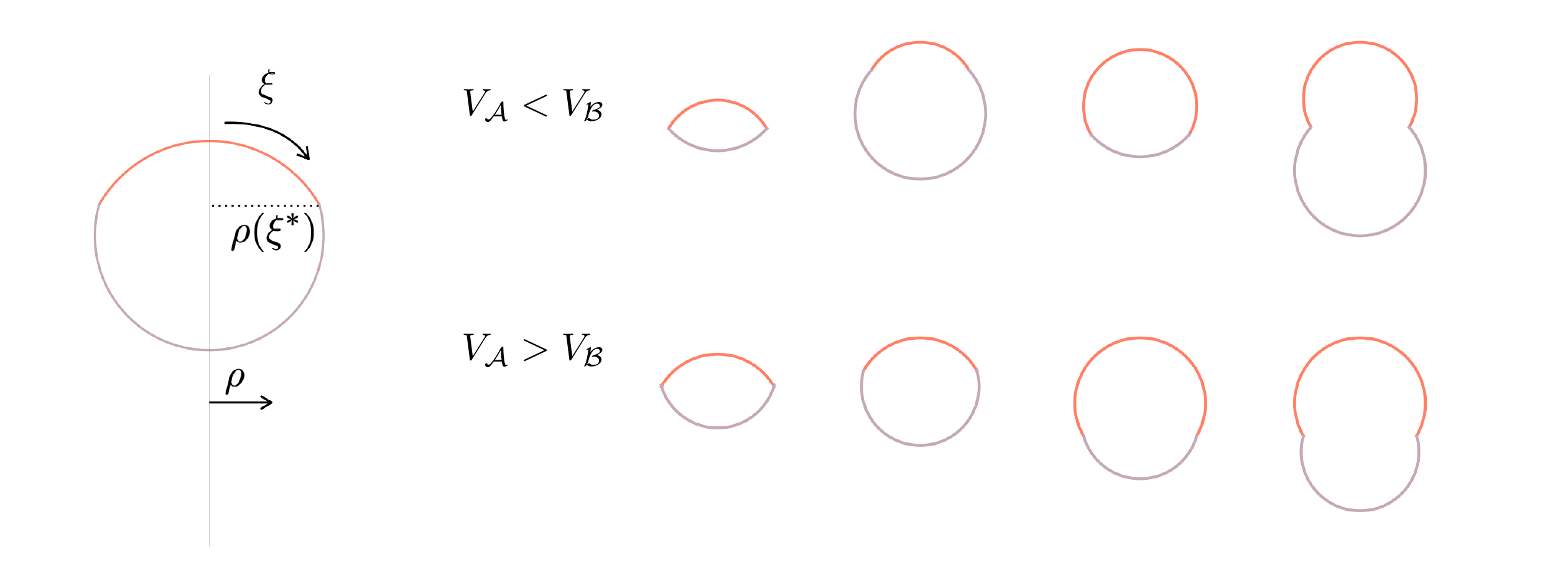}
\caption{\small The four different thin-wall configurations that can be constructed for a pair of dS vacua, red lines are for the daughter patch $\mathcal{B}$, gray are for the parent region $\mathcal{A}$. The top row would describe up-tunneling  (left to right: cap-to-cap, cap-to-bulb, bulb-to-cap, bulb-to-bulb). The bottom row would down-tunneling possibilities (same order). They are not drawn keeping $\rho(\xi^\ast)$  necessarily constant, as at this point they merely represent possibilities for dS-dS bounces. }
 \label{fig-dSdS-all}
\end{figure}

The Euclidean action for a thin-wall configuration \eqref{eq-phi-pcwise}-\eqref{eq-rho-pcwise} consists of three pieces --- the vacuum contributions from the regions on either side of the wall, plus the contribution from the wall. For a given parent-daughter pair of dS vacua, there are four different types of thin-wall configurations possible. We can connect a cap of the daughter to a bulb of the parent, a bulb of the daughter to a cap of the parent, two caps, or two bulbs. Parameterized in terms of the radius of the 3-sphere at juncture, $\rho(\xi^\ast=R )$, their Euclidean actions differ from the homogeneous parent configuration by\footnote{We use the convention that the word preceding ``to" in a hyphenated name, ``X-to-Y", always describes the region of new vacuum, $\xi<\xi^\ast$, while the word following ``to" refers to the region of unchanged vacuum, $\xi>\xi^\ast$. For example, a bulb-to-cap configuration consists of a bulb of $V_{\mathcal{B}}$ connected to a cap of $V_{\mathcal{A}}$.},
\begin{align}
B_{\text{cap-to-bulb}}&=F_1(V_{\mathcal{B}},R)+F_2(V_{\mathcal{A}},R) +B_{\text{wall}} - (F_1(V_{\mathcal{A}},R)+F_2(V_{\mathcal{A}},R))\nonumber\\
&=F_1(V_{\mathcal{B}},R)-F_1(V_{\mathcal{A}},R)+B_{\text{wall}}.\label{eq-B-capbulb}\\
B_{\text{cap-to-cap}}&=F_1(V_{\mathcal{B}},R)-F_2(V_{\mathcal{A}},R)+B_{\text{wall}}\nonumber\\
B_{\text{bulb-to-cap}}&=F_2(V_{\mathcal{B}},R)-F_2(V_{\mathcal{A}},R)+B_{\text{wall}}\nonumber\\
B_{\text{bulb-to-bulb}}&=F_2(V_{\mathcal{B}},R)-F_1(V_{\mathcal{A}},R)+B_{\text{wall}}.\nonumber
\end{align}
where $B_{\text{wall}}=2\pi^2 \sigma R^3$ and $\sigma$ is the tension of the thin wall (defined in terms of the 1-dimensional soliton action of an associated degenerate version of the theory). Using $F_2(V,R)=-\frac{24 \pi^2}{\kappa^2 V}-F_1(V,R)$ results in,
\begin{align}
B_{\text{cap-to-cap}}&=F_1(V_{\mathcal{B}},R)-(S_{\text{E}}[\mathcal{A}]-F_1(V_{\mathcal{A}},R)) +B_{\text{wall}}\nonumber\\
&=F_1(V_{\mathcal{B}},R)+F_1(V_{\mathcal{A}},R)+B_{\text{wall}}-S_{\text{E}}[\mathcal{A}]\label{eq-B-capcap}\\
B_{\text{bulb-to-cap}}&=(S_{\text{E}}[\mathcal{B}]-F_1(V_{\mathcal{B}},R))-(S_{\text{E}}[\mathcal{A}]-F_1(V_{\mathcal{A}},R))+B_{\text{wall}}\nonumber\\
& =F_1(V_{\mathcal{A}},R)-F_1(V_{\mathcal{B}},R)+B_{\text{wall}}+S_{\text{E}}[\mathcal{B}]-S_{\text{E}}[\mathcal{A}]\label{eq-B-bulbcap}\\
B_{\text{bulb-to-bulb}}&=(S_{\text{E}}[\mathcal{B}]-F_1(V_{\mathcal{B}},R))-F_1(V_{\mathcal{A}},R)+B_{\text{wall}}\nonumber\\
&= -F_1(V_{\mathcal{A}},R)-F_1(V_{\mathcal{B}},R)+S_{\text{E}}[\mathcal{B}]+B_{\text{wall}}.\label{eq-B-bulbbulb}
\end{align}
The Euclidean spacetime geometries for each of these four types are shown schematically in figure \ref{fig-dSdS-all}.

To see which of these piece-wise geometries can approximate extrema of $S_{\text{E}}$, we need to determine whether $\partial B/\partial R$ vanishes at any nontrivial $R$ within the physical domain, $R\in (0,\text{Min}(\ell_{\mathcal{A}}, \ell_{\mathcal{B}}))$. The cap function $F_1$ is a monotonically decreasing function of $R$,
\be
F_1=\frac{c_1}{V}\left( f(V,R)^3-1\right)
\ee
where the constant $c_1=\frac{12\pi^2}{\kappa^2}$, and recall $f(V,R)\equiv\sqrt{1-\kappa V R^2/3}$. For positive values of $V$, $f$ is a decreasing function of $R$, making $F_1$ a decreasing function as well. (For negative values of $V$, $f$ is an increasing function, but the sign of the overall factor $\frac{c_1}{V}$ is negative, making $F_1$ again a decreasing function. This will be useful in section \ref{sec-AdS}.) 

This immediately eliminates the bulb-to-bulb geometry because its decay exponent is the sum of three monotonically increasing functions: $ -F_1(V_{\mathcal{A}},x)$, $-F_1(V_{\mathcal{B}},x)$ and the cubic, $B_{\text{wall}}$, (plus the constant $S_{\text{E}}[\mathcal{B}]$, which is irrelevant to this argument), and so it is not possible to have vanishing $B'$. Solely based on the monotonicity of $F_1$ and $B_{\text{wall}}$, the remaining three geometries cannot be eliminated because each of their tunneling exponents involve a combination of at least one monotonically increasing and one monotonically decreasing function.

Proceed by considering a difference in cap functions, $F_1(V,x)-F_1(\tilde{V},x)$. This is a monotonically decreasing function of $x$ for $V<\tilde{V}$, and a monotonically increasing one for $V>\tilde{V}$. This can been seen from its derivative with respect to $x$,
\be
\frac{\partial (F_1(V,x)-F_1(\tilde{V},x))}{\partial x}=c_1 \kappa x(f(\tilde{V},x)-f(V,x)).\label{diff-cap-functions-deriv}
\ee
Apart from the origin where the two  $f$ square root functions coincide, the difference $f(\tilde{V},x)-f(V,x)$ is strictly negative if $\tilde{V}>V$, and strictly positive if $V>\tilde{V}$. 
The negativity of $f(\tilde{V},x)-f(V,x)$  when $\tilde{V}>V$ implies $F_1(V,x)-F_1(\tilde{V},x)$ decreases monotonically over $[0,\tilde{\ell}]$ because \eqref{diff-cap-functions-deriv} is negative. If instead $\tilde{V}<V$ the same reasoning implies $F_1(V,x)-F_1(\tilde{V},x)$ is monotonically increasing, though notice the physical interval is now defined as $[0,\ell]$. At the origin, $x=0$, the slope of $F_1(V,x)-F_1(\tilde{V},x)$ is zero.

Due to the monotonically increasing contribution, $B_{\text{wall}}$, the cap-to-bulb geometry must have $F_1(V_{\mathcal{B}},x)-F_1(V_{\mathcal{A}},x)$  \emph{decreasing} in order for $B$ is to have an extremum. Given our observation about the difference in two cap functions, $V_{\mathcal{A}}<V_{\mathcal{B}}$ decays (i.e. uptunneling) are ruled out for the cap-to-bulb geometry. The bulb-to-cap case, on the other hand, involves the opposite difference, $F_1(V_{\mathcal{A}},x)-F_1(V_{\mathcal{B}},x)$ and thus \emph{downtunneling} is ruled out for this bubble geometry. We shall shortly see that these describe the same bubble, as expected, only with the ``inside" and ``outside" labels assigned in the opposite manner. The cap-to-cap geometry instead involves the sum of two cap functions, as opposed to their difference. We'll return to this case after we finish with the analysis for cap-to-bulb downtunneling, and bulb-to-cap uptunneling.

The derivatives of $B$ for these work out to,  
\begin{align}
B'_{\text{cap-to-bulb}}&=c_1 \kappa x ( f( V_{\mathcal{A}},x)-f( V_{\mathcal{B}},x)+\sigma \kappa x/2)\label{Bprime-cap-bulb}\\
B'_{\text{bulb-to-cap}}&= c_1  \kappa x ( f( V_{\mathcal{B}},x)-f( V_{\mathcal{A}},x)+\sigma \kappa x/2)\label{Bprime-bulb-cap}.
\end{align}
To take into account the restriction --- downtunneling for cap-to-bulb, and uptunneling for bulb-to-cap ---  we'll express things in terms the lower and higher dS vacuum energies,  $V_0$ and $V_0+\Delta V$, where $\Delta V>0$. The restriction means making the assignment  $V_{\mathcal{B}}=V_0$ and $V_{\mathcal{A}}=V_0+\Delta V$ for cap-to-bulb, and the opposite assignment for bulb-to-cap, $V_{\mathcal{B}}=V_0+\Delta V$ and $V_{\mathcal{A}}=V_0$. With this notation we see that the $B'$ for these two processes --- cap-to-bulb downtunneling and bulb-to-cap uptunneling --- are the same, 
\be
B'_{\text{cap-to-bulb, down}}=B'_{\text{bulb-to-cap, up}}=c_1 \kappa x ( \Delta f+\sigma \kappa x/2)\label{B'-expr1}
\ee
where $\Delta f\equiv f( V_0+\Delta V,x)-f( V_0,x)$. Thus, their $B$'s are extremized at the same $R$ values, making their patched configurations identical under reflection of $\xi$ across $\xi^\ast$. The \emph{values} of $B$, of course, are not the same. They differ, by construction, by $S_{\text{E}} [V_0 ]-S_{\text{E}}[ V_0+\Delta V ]$; in line with detailed balance\footnote{This can be seen by comparing expressions \eqref{eq-B-capbulb} and \eqref{eq-B-bulbcap}.
\begin{align*}
B_{\text{cap-to-bulb, down}}-B_{\text{bulb-to-cap, up}}=&F_1(V_0,R)-F_1(V_0+\Delta V,R)+B_{\text{wall}}\\
&-\left( F_1(V_0,R)-F_1(V_0+\Delta V,R)+B_{\text{wall}}+S_{\text{E}}[V_0+\Delta V]-S_{\text{E}}[V_0]\right) \\
&=S_{\text{E}}[V_0]- S_{\text{E}}[V_0+\Delta V]
\end{align*} 
By the shorthand  $S_{\text{E}}[V]$ we mean the Euclidean action of a full dS 4-sphere for vacuum energy density $V$.}.

Consider the Taylor expansion of $f(V,x)$ near the origin,
\be
f(V,x)=1-\frac{\kappa V}{6}x^2+\mathcal{O}(x^3),
\ee
and so $\Delta f\approx -\frac{\kappa\Delta V}{6}x^2$. The positive linear term $\sigma \kappa x/2$ in \eqref{B'-expr1} coming from the wall contribution therefore dominates $\Delta f$ near the origin. With the overall factor of $x$ in \eqref{B'-expr1}, this causes $B'$ to be an increasing function from zero near the origin, 
$$B'\approx \frac{1}{2}c_1 \kappa^2 \sigma x^2.$$
For there to exist an extremum at nontrivial $x$, $\Delta f$ needs to become more negative than $-\sigma \kappa x/2$ somewhere in the physical interval. 
Proceed by pulling out the factor of $\kappa \sigma x/2$ in $B'$,
\be
B'=c_2 x^2 \left( \frac{\Delta f}{\kappa \sigma x/2}+1\right)\label{B'-expr2}
\ee
with $c_2\equiv \kappa^2 \sigma c_1/2=6\pi^2 \sigma$. Define the function $Z(x) \equiv  \frac{\Delta f}{\kappa \sigma x/2}$, parameterized by $V_0$, $\Delta V$, and $\sigma$.

$Z$ is monotonically decreasing with $x$, and negative everywhere aside from the origin, where $Z=0$ (recall $\Delta f$ is a negative quadratic there). Hence, $B$ has at most  one nontrivial extremum. It exists if $Z<-1$ anywhere in the interval $[0,\ell_+]$, where $\ell_+$ is the smaller of the dS radii: $\ell_+=\sqrt{3/(\kappa (V_0+\Delta V))}$. This extremum --- if it exists --- is necessarily a local maximum due to the monotonicity of $Z$, and the fact that $B'>0$ for small, but nonzero $x$. In other words, if $B'$ crosses zero at some $R>0$, it changes sign from positive to negative, signifying the presence of a local maximum of $B$. The existence of the local maximum can be guaranteed by simply evaluating $Z(x=\ell_+)$ and seeing if it is less than $-1$. 

That $Z$ is monotonically decreasing can be seen from the fact that $\partial Z/\partial x$,
\begin{align}
\frac{\partial Z}{\partial x}&=\frac{2}{\sigma \kappa}\left[\left(\frac{V_0 \kappa/3}{f_0}+\frac{f_0}{x^2}\right)-\left(\frac{V_+ \kappa/3}{f_+}+\frac{f_+}{x^2}\right)\right]\label{Z-prime-1}\\
&=\frac{2}{\sigma \kappa x^2}\left[\left(\frac{x^2 V_0 \kappa/3+f_0^2}{f_0}\right)-\left(\frac{x^2 V_+ \kappa/3+f_+^2}{f_+}\right)\right]\label{Z-prime-3}\\
&=\frac{2}{\sigma \kappa x^2}\left(\frac{1}{f_0}-\frac{1}{f_+} \right)\label{Z-prime-4},
\end{align}
 is negative everywhere in the physical interval\footnote{$Z'$ is finite at the origin because $(f_0-f_+)/x^2=-\Delta f/x^2\sim +\kappa |\Delta V| /6$ is finite. Indeed $Z'<0$ here because: $$Z'(0)=\frac{2}{\sigma \kappa}\left[\frac{1}{\ell_0^2}-\frac{1}{\ell_+^2}+\frac{\kappa \Delta V}{6}\right]= \frac{2}{\sigma \kappa}\left[\frac{\kappa V_0}{3}-\frac{\kappa (V_0+\Delta V)}{3} +\frac{\kappa \Delta V}{6} \right]=\frac{2}{\sigma \kappa}\left( -\frac{\kappa \Delta V}{6}\right)<0$$}.
This follows from the fact that $f_0=\sqrt{1-x^2 V_0 \kappa/3}$ is \emph{larger} than $f_+=\sqrt{1-x^2 V_+ \kappa/3}$ because $V_0<V_+=V_0+\Delta V$. Thus, $\frac{1}{f_0}<\frac{1}{f_+}$, making $\partial Z/\partial x<0$ by equation \eqref{Z-prime-4}. Therefore, $Z$ is monotonically decreasing over the physical interval, as claimed.

Now we'll find the most negative value of $Z$ by evaluating at $x=\ell_+$. Here $\Delta f=0-\sqrt{1-\ell_+^2/\ell_0^2}$, so
\be 
Z(\ell_+)=-\frac{\sqrt{1-\ell_+^2/\ell_0^2}}{\kappa \sigma \ell_+/2}.\label{minZ}
\ee  
This is only less than $-1$ if,
\be
\sigma<\frac{2}{\kappa}\sqrt{\frac{1}{\ell_+^2}-\frac{1}{\ell_0^2}}.\label{ineq-cap-bulb}
\ee
Since the right-hand side depends only on the vacuum energies, we recognize that for each pair of dS vacua there is a threshold value for the tension $\sigma$ above which instantons with the cap-to-bulb/bulb-to-cap geometry cease existing. 

 \begin{table}[h]
  \begin{center}
    \caption{\small Summary of dS-dS bounces.}
    \label{tab:table1}
    \begin{tabular}{|l c | c | c | c | } 
    \hline
      \textbf{geometry} & & $\sigma< \frac{2}{\kappa}\sqrt{\frac{1}{\ell_+^2}-\frac{1}{\ell_0^2}}$ & $\sigma> \frac{2}{\kappa}\sqrt{\frac{1}{\ell_+^2}-\frac{1}{\ell_0^2}}$& for \includegraphics[width=.02\linewidth]{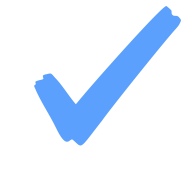} varying $R$ decreases $S_{\text{E}}$? \\
      \hline
      \textbf{cap-to-bulb} & & & & \\
   $V_{\mathcal{B}}>V_{\mathcal{A}}$ 
  &  \includegraphics[width=.08\linewidth]{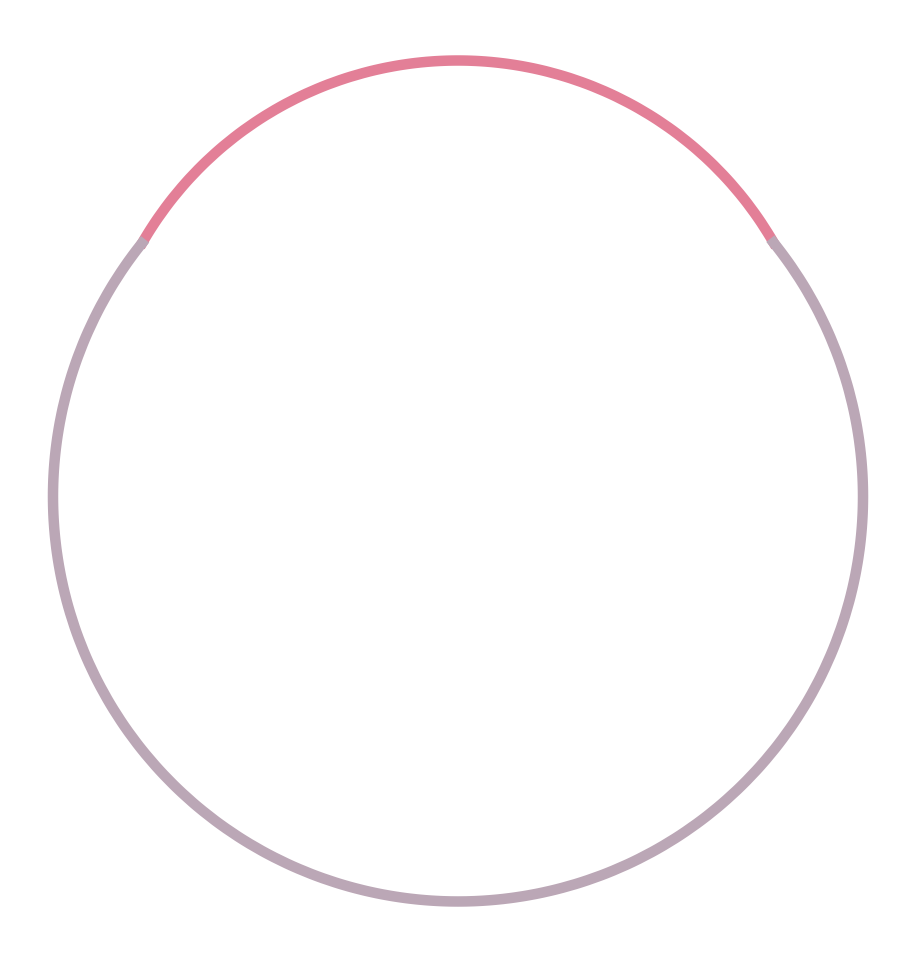} &  \includegraphics[width=.04\linewidth]{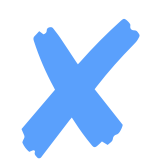} & \includegraphics[width=.04\linewidth]{for_table_x.png}   & N/A\\
   & & & & \\
        $V_{\mathcal{A}}>V_{\mathcal{B}}$ &\includegraphics[width=.08\linewidth]{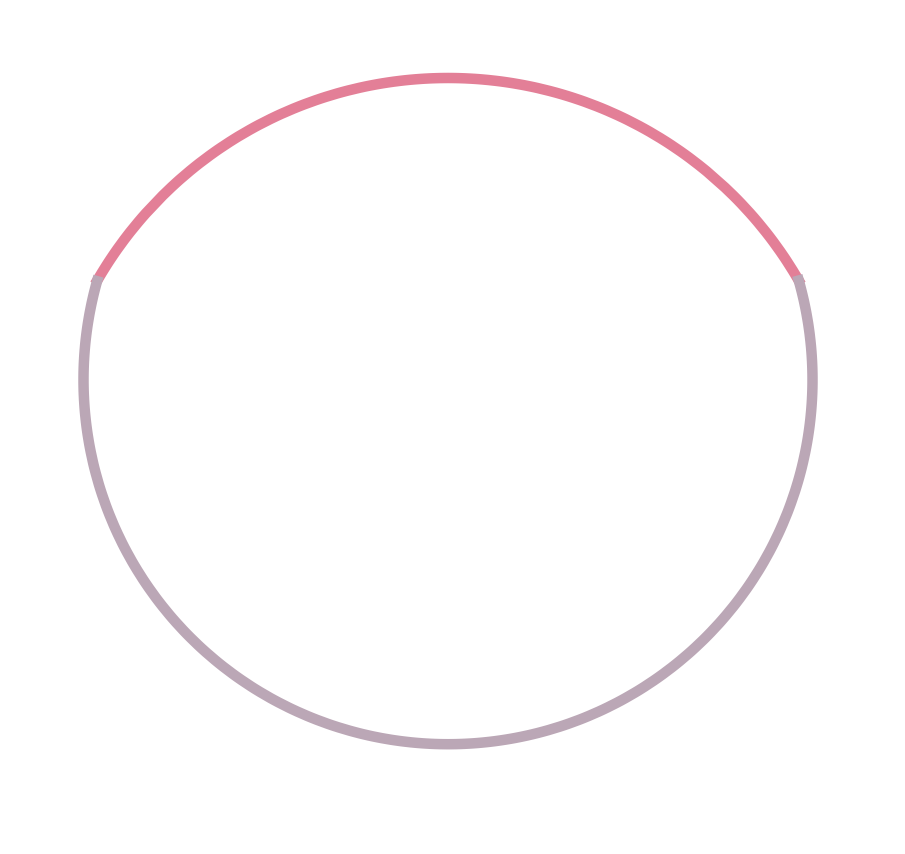} &  \includegraphics[width=.05\linewidth]{for_table_check.png} &  \includegraphics[width=.04\linewidth]{for_table_x.png} & yes     \\
 &  & & &  \\
      \hline
    \textbf{bulb-to-cap}  &   & & & \\
      $V_{\mathcal{B}}>V_{\mathcal{A}}$  &  \includegraphics[width=.08\linewidth]{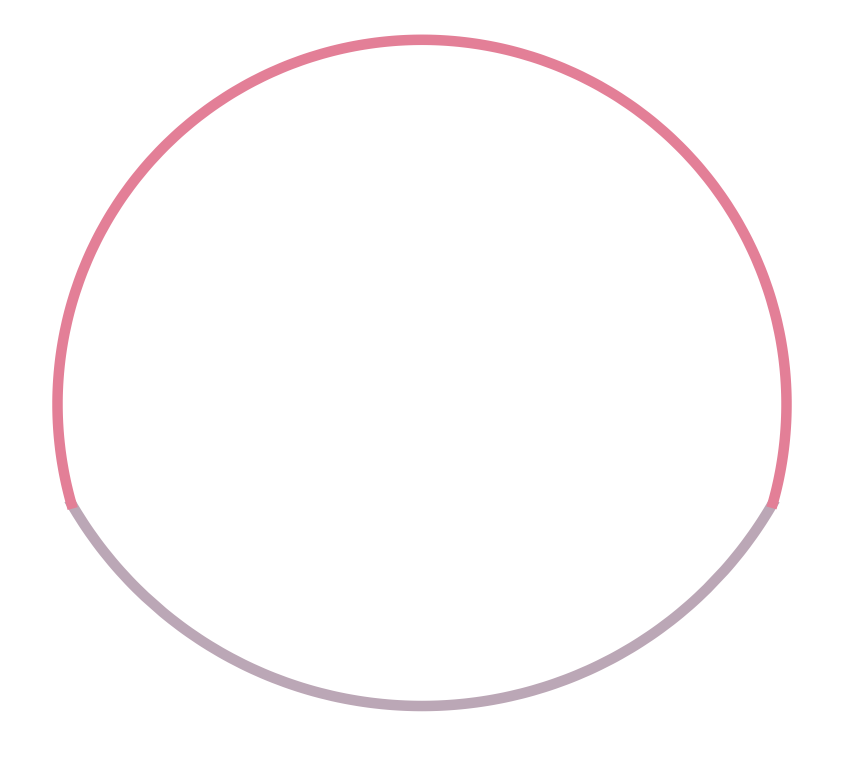} &  \includegraphics[width=.05\linewidth]{for_table_check.png} & \includegraphics[width=.04\linewidth]{for_table_x.png} &   yes \\
   && & & \\
       $V_{\mathcal{A}}>V_{\mathcal{B}}$ &\includegraphics[width=.08\linewidth]{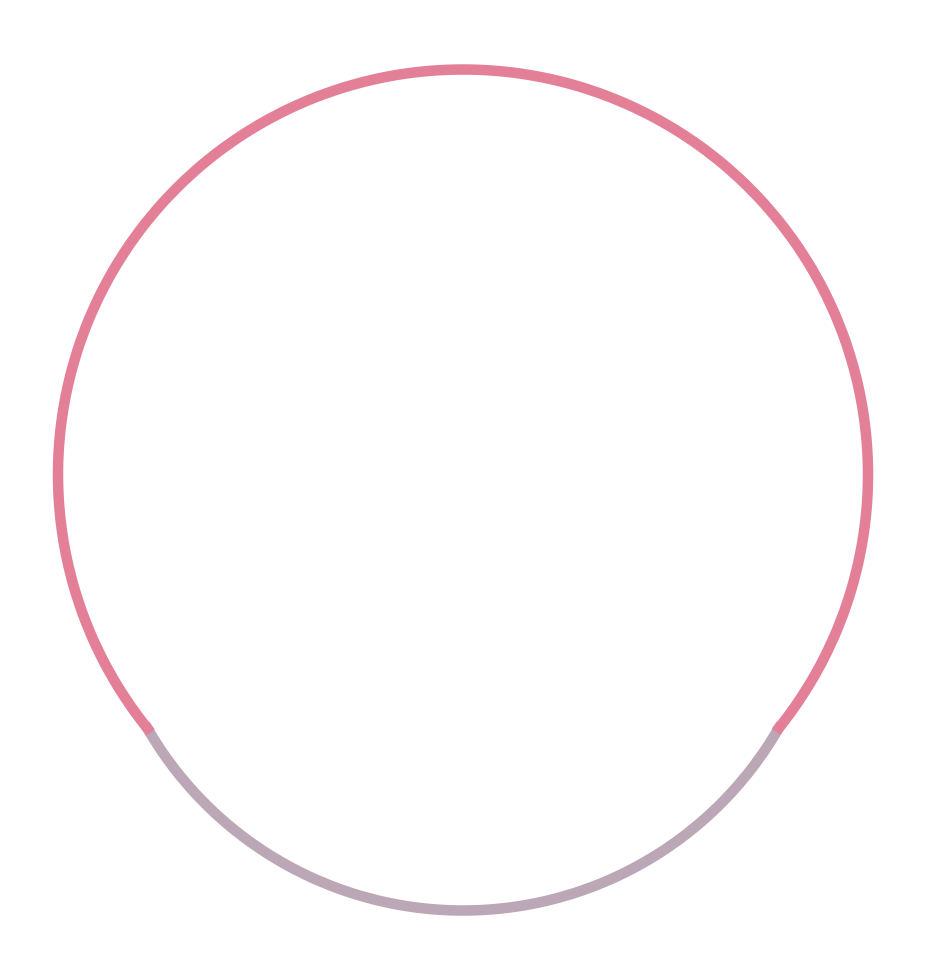} &  \includegraphics[width=.04\linewidth]{for_table_x.png} &  \includegraphics[width=.04\linewidth]{for_table_x.png} &  N/A  \\
 &  && & \\
\hline
        \textbf{cap-to-cap}&  & & & \\
            $V_{\mathcal{B}}>V_{\mathcal{A}}$  &  \includegraphics[width=.08\linewidth]{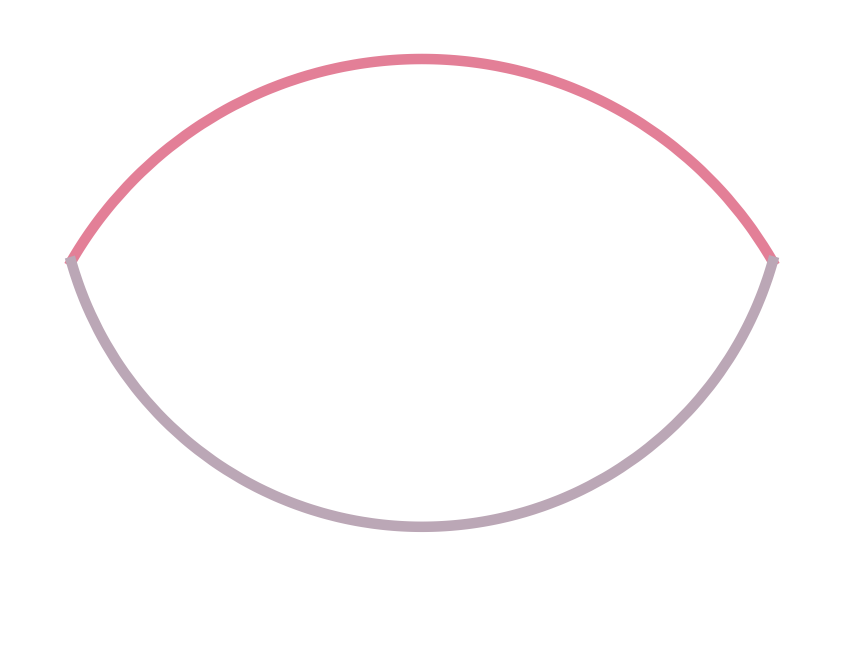} &   \includegraphics[width=.04\linewidth]{for_table_x.png} &   \includegraphics[width=.05\linewidth]{for_table_check.png}& no \\
   && & & \\
      $V_{\mathcal{A}}>V_{\mathcal{B}}$  &\includegraphics[width=.08\linewidth]{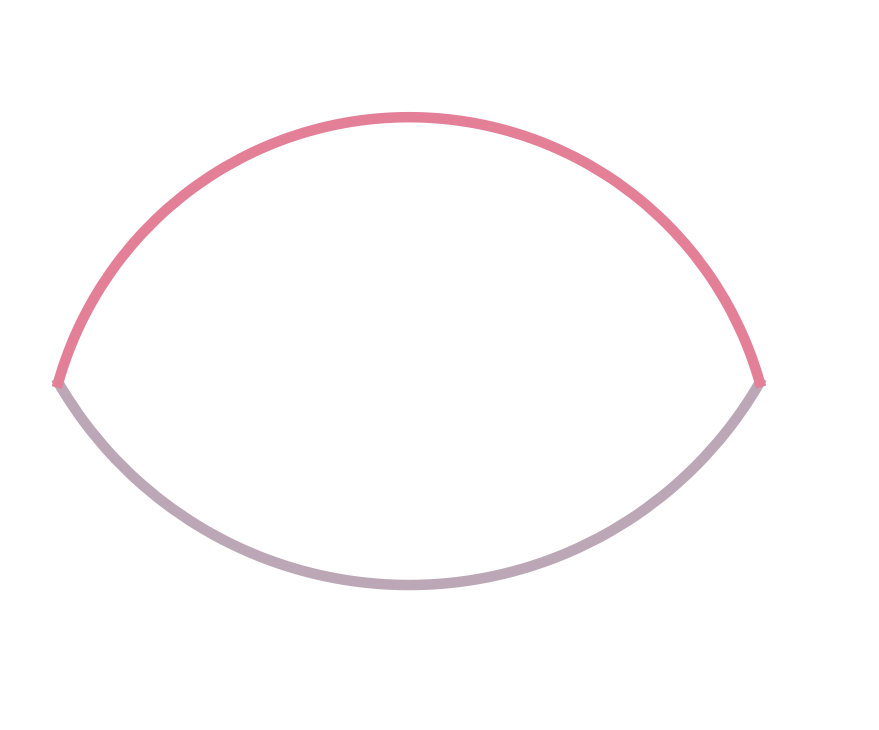} &   \includegraphics[width=.04\linewidth]{for_table_x.png} & \includegraphics[width=.05\linewidth]{for_table_check.png} &  no  \\
 &  && &\\
\hline
  \textbf{bulb-to-bulb}& &&&\\
   $V_{\mathcal{B}}>V_{\mathcal{A}}$  &  \includegraphics[width=.08\linewidth]{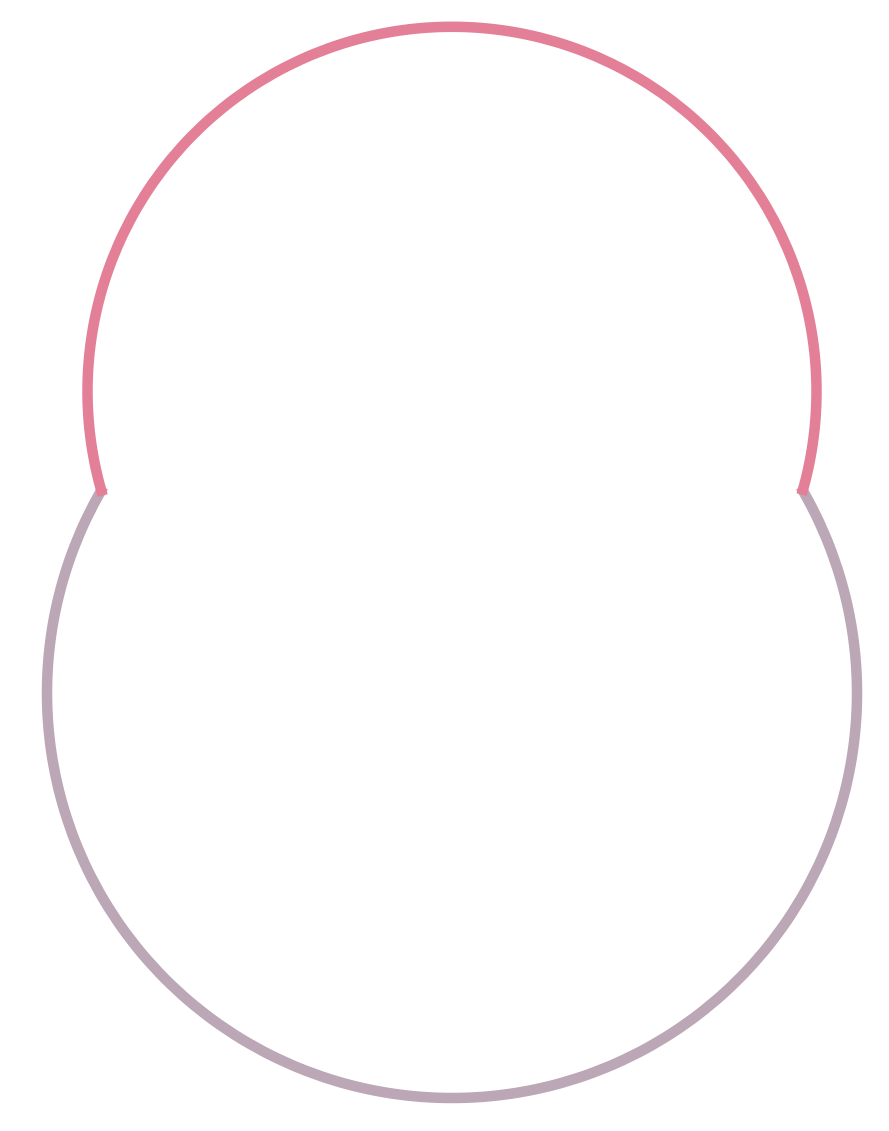} &  \includegraphics[width=.04\linewidth]{for_table_x.png} & \includegraphics[width=.04\linewidth]{for_table_x.png}   & N/A   \\
  && & & \\
    $V_{\mathcal{A}}>V_{\mathcal{B}}$ &\includegraphics[width=.08\linewidth]{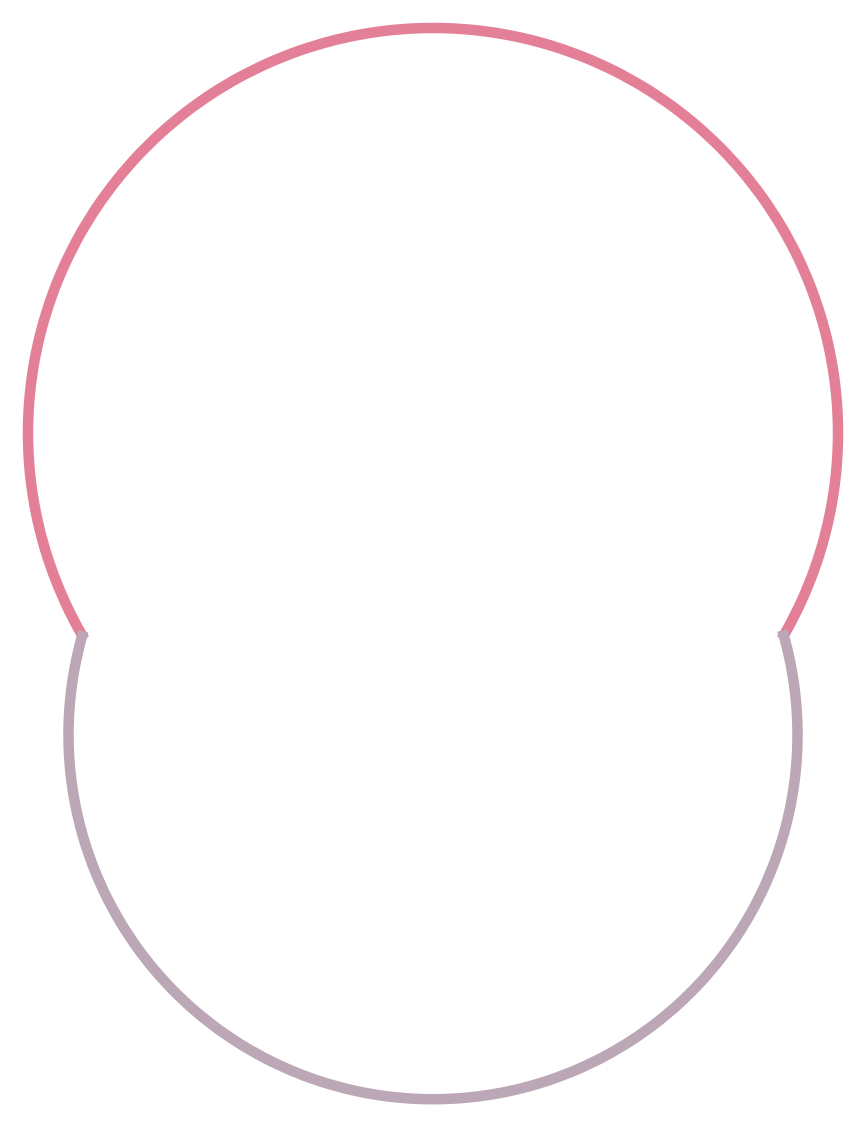} &  \includegraphics[width=.04\linewidth]{for_table_x.png} &  \includegraphics[width=.04\linewidth]{for_table_x.png} &  N/A  \\
 &  && &\\
             \hline
    \end{tabular}
  \end{center}
\end{table}

Next we turn to the cap-to-cap geometry. Its $B'$ involves the \emph{sum} of two $f$ functions, instead of their difference $\Delta f$, because $B_{\text{cap-to-cap}}$ involves the sum of parent and daughter $F_1$'s, as opposed to their difference. Expressed in terms of the average $\bar{f}(x)\equiv \frac{f(V_0,x)+f(V_0+\Delta V, x)}{2}$, we have
\be
B'_{\text{cap-to-cap}}=c_1 \kappa x\left(-2\bar{f}(x) +\sigma \kappa x/2\right)\label{B'-capcap1}.
\ee
This holds for both up and downtunneling because the sum of the $F_1$'s is symmetric under $\mathcal{A}\leftrightarrow\mathcal{B}$. Note that $B_{\text{cap-to-cap, up}}$ and $B_{\text{cap-to-cap, down}}$ again differ, by construction, by $S_{\text{E}}[V_0]-S_{\text{E}}[V_0+\Delta V]$. 

The function $-2\bar{f}(x)$ is monotonically increasing from $-2$ at $x=0$ because $f$ functions are monotonically decreasing from $+1$ for positive vacuum energies. $-2\bar{f}(x)$ therefore dominates the linear wall term $\sigma \kappa x/2$ in \eqref{B'-capcap1} near the origin, making $B'$ negative for small, but nonzero $x$. Notice that the extremum of $B$ exactly at the origin, which is a consequence of $B'$ in \eqref{B'-capcap1} containing an overall factor of $x$, corresponds to a local maximum of $B$ because,
\begin{align*}
&B'_{\text{cap-to-cap}}\approx c_1 \kappa x\left(-2(1+\mathcal{O}(x^2)) + \sigma \kappa x/2\right)\\
&B''_{\text{cap-to-cap}|_{x=0}}=-2c_1 \kappa <0.
\end{align*}
As we move to the right from zero, $B(x)$ decreases (we are on the side of a hill). We will show that aside from the origin there is at most one additional turning point of $B$, a local minimum. 

Due to the monotonicity of $-2\bar{f}(x)$, the term in parenthesis in \eqref{B'-capcap1} is a monotonically increasing function that starts off with a negative value. For this reason $B'$ will either eventually become zero again, after which it continues to increase taking only positive values, or we will reach the size of the smaller dS radius. Consequently, the local maximum in $B$ at the origin is either followed by a local minimum ($B'$ changes from negative to positive across the extremum), or no extremum at all. 
Hence it suffices to evaluate $-2\bar{f}$ where it takes its largest value --- at the maximum value of $x$ in the physical domain, $\ell_+$ --- and see whether it is large enough to make $B'>0$. That is, impose
\be
-2\bar{f}(\ell_+)>-\sigma \kappa \ell_+/2.
\ee 
Since $2\bar{f}(\ell_+)=f_0(\ell_+)=\sqrt{1-\ell_+^2/\ell_0^2}$,  we find the condition, 
\begin{align}
\sigma&> \frac{2}{\kappa}\sqrt{\frac{1}{\ell_+^2}-\frac{1}{\ell_0^2}}\label{ineq-cap-cap}.
\end{align}

Inequality \eqref{ineq-cap-cap} is precisely the reverse of the condition we found for the instanton involving a cap connected to a bulb to exist, \eqref{ineq-cap-bulb}. This indicates that for any pair of dS vacua and value of the tension, there exists a spherically symmetric thin-wall configuration interpolating between the two whose wall follows an $SO(3,1)$-invariant trajectory in the Lorentzian theory. If the tension is below the threshold,
\be
\frac{2}{\kappa}\sqrt{\frac{1}{\ell_+^2}-\frac{1}{\ell_0^2}}
\ee
the configuration corresponds to an instanton that has the traditional expansion/contraction negative mode in the Euclidean theory, whereas if the tension is above the threshold the configuration corresponds to an extremum of $S_{\text{E}}$ that does not have this negative mode. The former are referred to in the literature as type-A instantons, while the latter are referred to as type-B. Our results are summarized in table \ref{tab:table1}.
\begin{figure}[h!]
  \centering
  \includegraphics[width=1\linewidth]{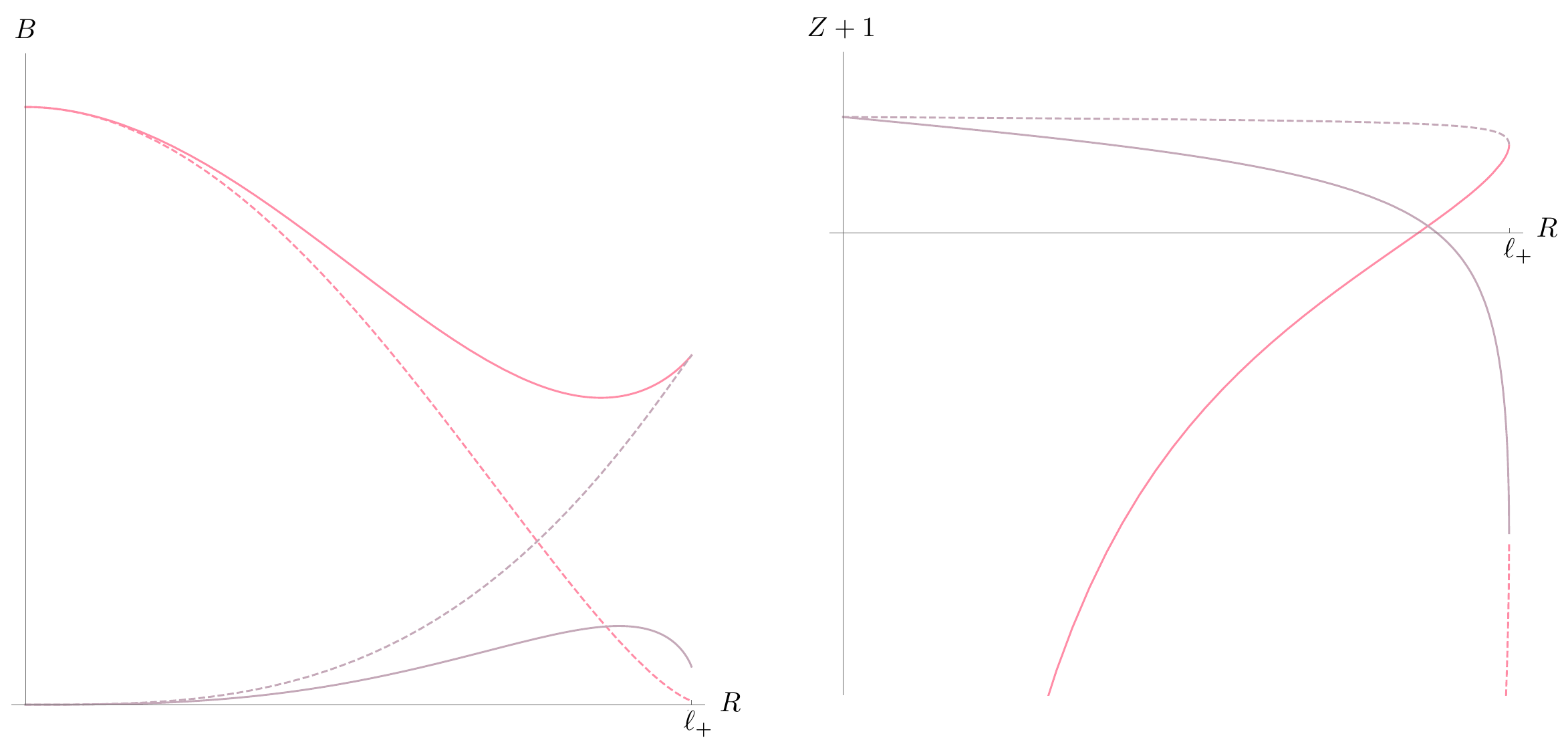}
  \caption{\small For a fixed pair of dS energies $V_0$, and $V_0+\Delta V$, plots of the cap-to-bulb $B$ (left) and $Z+1$ (right) are shown in gray for two choices of the tension, $\sigma_1$ and $\sigma_2$. The solid gray lines are for $\sigma_1$, which satisfies inequality \ref{ineq-cap-bulb}, while the dashed gray lines are for $\sigma_2$, which does not. Shown in pink on the same sets of axes are the analogous quantities for the cap-to-cap bounce, for the same $\sigma$ values. We have used solid pink lines for the cap-to-cap bounce with $\sigma_2$ (which satisfies its inequality, \ref{ineq-cap-cap} because it is the reverse of \ref{ineq-cap-bulb}). The cap-to-cap functions using $\sigma_1$ are shown in dashed pink. By $Z$ for cap-to-cap we mean the function $\frac{-2\bar{f}}{\kappa \sigma R/2}$. Finally, for the cap-to-bulb instanton with $\sigma_1$ we have scaled $B$ by a factor of $10$ so that the maximum is easier to see.}. 
  \label{fig-capbulb-plots}
  \end{figure}
\begin{figure}
 \centering
\includegraphics[width=.4\linewidth]{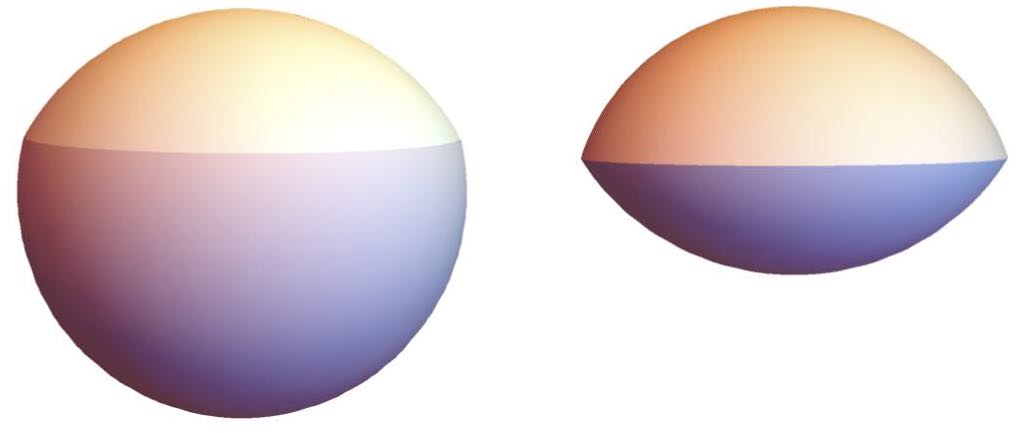}
\caption{\small Representations of the type-A bounce (left), and type-B bounce (right) as 2-manifolds embedded in $\mathbb{R}^3$. Type-A involves a cap and bulb, whereas type-B involves two caps. The $\xi$ coordinate runs longitudinally around each manifold. For the downtunneling transitions, the centers of the bounces ($\xi=0$) are located at the centers of the northern caps. For uptunneling, they lie is at the centers of the southern patches. Surfaces of constant $\xi$ are circles drawn latitudinally (the height of the horizontal plane containing the circle is indicative of $\xi$). The radius of such a circle is $\rho(\xi)$, and each circle should be interpreted as a 3-sphere.}
 \label{cap-to-bulb-3d}
\end{figure}

\section{Decays to AdS}\label{sec-AdS}
\subsection{AdS-AdS transitions}
First we consider the case where both the parent and daughter vacua are AdS. Labeling the parent by the $\mathcal{A}$ subscript as usual, begin with
\be 
B_{\text{AdS-AdS}}= S_{\text{E}}[\text{inst}]-S_{\text{E}}[ \mathcal{A}].\label{B-general-AdS-sec}
\ee
The $\xi$ coordinate-range for AdS is $[0,\infty)$. The parent action diverges like, 
$$
S_{\text{E}}[ \mathcal{A}]=\lim_{\xi \rightarrow\infty}  - \frac{4\pi^2\ell^2}{\kappa} \left(\cosh^3(\xi/\ell)-1\right)=-\infty.
 $$

Despite the infinite AdS action, $B$ may nevertheless be finite because the divergence can be canceled by the contribution to the instanton action from the $\xi>\xi^\ast$ region,  where  $\rho_{\text{inst}}$ is exponentially close to $\rho_\mathcal{A}$. Specifically, only a 4-geometry consisting of a finite patch of the $\mathcal{B}$ AdS  connected to the \emph{complement} of the analogous finite patch for the $\mathcal{A}$ AdS  results in a finite $B$. It can be computed as follows. Denoting the $\rho$ value at the position of the wall by $x$, \be 
S_{\text{E}}[ \mathcal{A}]= F_1(V_{\mathcal{A}},\infty)=F_1(V_{\mathcal{A}},x)+4\pi^2 \int_{\rho^{-1}_{\mathcal{A}}(x)}^{\infty} d\xi  \left(\rho_{\mathcal{A}}^3 V_{\mathcal{A}}-\frac{3\rho_{\mathcal{A}}}{\kappa}\right)
\ee
and 
\be 
S_{\text{E}}[\text{inst}]= F_1(V_{\mathcal{B}},x)+ B_{\text{wall}}+4\pi^2 \int_{\rho^{-1}_{\mathcal{A}}(x)}^{\infty} d\xi  \left(\rho_{\mathcal{A}}^3   V_{\mathcal{A}}-\frac{3\rho_{\mathcal{A}}}{\kappa}\right),
\ee
so the difference is,
\be 
B_{\text{patch-to-comp}}= F_1(V_{\mathcal{B}},x)-F_1(V_{\mathcal{A}},x)+B_{\text{wall}}.\label{eq-B-patch-comp}
\ee

The form of \eqref{eq-B-patch-comp} is identical to the cap-to-bulb case for dS-dS instantons. The only difference is that the energy densities $V_{\mathcal{A}}$ and $V_{\mathcal{B}}$ take negative values now. Just as it was for positive energies, the function $F_1(V_{\mathcal{B}},x)-F_1(V_{\mathcal{A}},x)$ is monotonic with $x$ for negative energies. The criteria determining whether $F_1(V_{\mathcal{B}},x)-F_1(V_{\mathcal{A}},x)$ is increasing or decreasing is the same: for $V_{\mathcal{B}}<V_{\mathcal{A}}$ the difference is monotonically decreasing, and for the reverse it is increasing\footnote{This can be seen from equation \eqref{diff-cap-functions-deriv}, which is valid regardless of the signs of $V$ and $\tilde{V}$. For negative energies, $f(V,x)=\sqrt{1+\kappa|V|x^2/3}$ increases with $x$ without bound starting from $f=1$ at the origin. The larger $|V|$ is the greater the rate at which it grows. So, for $\tilde{V}<V<0$, $f(\tilde{V},x)>f(V,x)$, making $f(\tilde{V},x)-f(V,x)$ positive at all nontrivial $x$, and $F_1(V,x)-F_1(\tilde{V},x)$  monotonically increasing. The same reasoning implies that when  $V<\tilde{V}<0$,  $F_1(V,x)-F_1(\tilde{V},x)$ is monotonically decreasing.}. Thus, uptunneling between the two AdS is ruled out ($B$ for such transitions is the sum of strictly monotonically increasing functions). 

The same factorization of $B'$ holds as in the cap-to-bulb/bulb-to-cap cases, $$B'_{\text{patch-to-comp, down}}=c_2 x^2 \left(Z+1\right)$$ with $Z$ and $\Delta f$ defined in the same manner, $$Z\equiv \frac{\Delta f}{\kappa \sigma x/2},$$ and $$\Delta f=f(V_0+\Delta V,x)-f(V_0,x).$$ All we need to do is analyze $Z$  for negative $V_0$ and $0<\Delta V<|V_0|$. 

Recall that when evaluated with negative energy, $f$ \emph{increases} monotonically with $x$. It starts from the value $f=1$ at the origin, just as in the dS case, but increases without bound over the now noncompact physical interval, $x\in [0,\infty)$. Near the origin, $f$ looks like a right-side up parabola, $f\approx 1+\frac{\kappa |V|}{6}x^2$. For large $x$, it is approximately linear, $f\sim \sqrt{\frac{\kappa |V|}{3}}x=x/\ell$. 

Although $f_0$ and $f_+$ are positive and increasing, their difference $\Delta f=f_+-f_0$ is still $\leq 0$  ($f_0=\sqrt{1+|V_0|\kappa x^2/3}$ increases more rapidly than $f_+=\sqrt{1+|V_0+\Delta V|\kappa x^2/3}$ because $|V_0+\Delta V|<|V_0|$). The Taylor expansion of $f$ again implies $Z(0)=0$, and the same algebraic manipulations of $\partial_x Z$  hold (\eqref{Z-prime-3} is valid for any value of $V_0$). We copy the result for convenience,
\be
\frac{\partial Z}{\partial x}=\frac{2}{\sigma \kappa x^2}\left(\frac{1}{f_0}-\frac{1}{f_+} \right).
\ee 
Since once again $f_+<f_0$, the difference of the inverses appearing in the parenthesis above is negative, making $Z+1$ a decreasing function over  $\mathbb{R}_+$ that takes positive values near the origin. At most one nontrivial root of $B'$ exists due to the monotonicity of $Z+1$. If $Z$ reaches $-1$ at some finite $x$, the sign of $B'$ changes from positive to negative across it, indicating that the extremum of $B$ is a local maximum. Therefore, patch-to-complement instantons describing downtunneling  possess the expansion/contraction negative mode.

The asymptotic form for $f$ implies that $Z$ approaches a constant, 
\be
Z\sim \frac{2}{\sigma \kappa x}  \left(\frac{x}{\ell_+}-\frac{x}{\ell_-} \right)= \frac{2}{\sigma \kappa }  \left(\frac{1}{\ell_+}-\frac{1}{\ell_-} \right)\equiv Z_{\infty}
\ee
Thus, the condition for the existence of the AdS-AdS thin-wall instanton, $Z_{\infty}<-1$, is, 
\be
\sigma < \frac{2}{\kappa} \left(\frac{1}{\ell_-}-\frac{1}{\ell_+} \right) \label{ineq-AdS}
\ee
where $\ell_+$ and $\ell_-$ are the radii of the higher and lower energy AdS, respectively,
\be
\ell_+=\sqrt{\frac{3}{\kappa (|V_0|-\Delta V)}}, \qquad  \ell_-=\sqrt{\frac{3}{\kappa |V_0|}}, \qquad \ell_+>\ell_-.
\ee

\subsection{dS-AdS transitions}
\begin{figure}
  \centering
  \includegraphics[width=.75\linewidth]{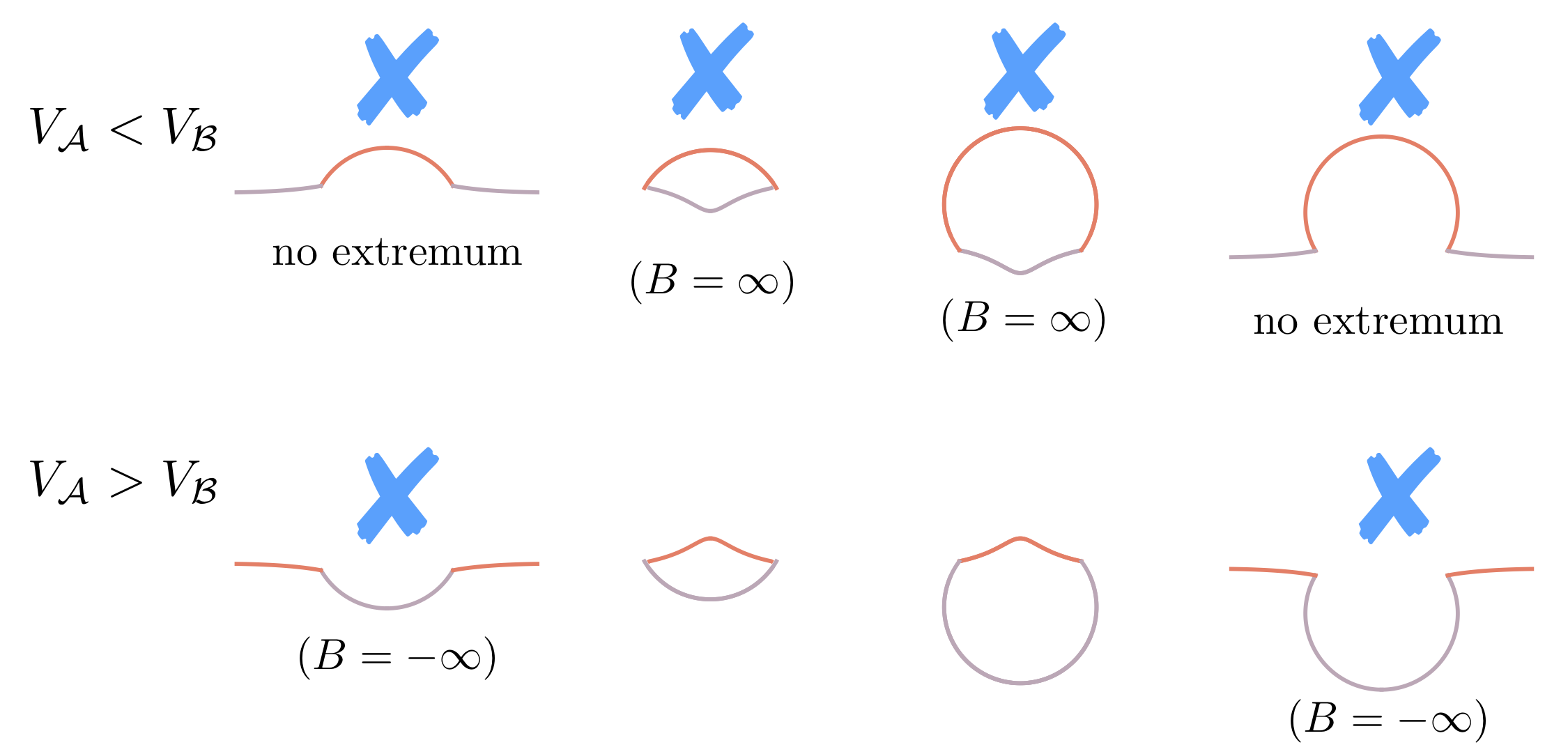}
  \caption{\small Possibilities for connecting dS and AdS regions. Same convention as figure \ref{fig-dSdS-all}: red for daughter vacuum configuration, gray for parent. All instantons potentially describing uptunneling (top row) are excluded either by failing to have finite $B$, or failing to extremize $B$. Viewing $B=-\infty$ as a nonsensical result, only two of the instatons potentially describing downtunneling (bottom row) are possible: patch-to-cap (second from the left) and patch-to-bulb (third from the left).}
\label{fig-mixed-cases}
\end{figure}

For mixed transitions --- one AdS and one dS --- there are four distinct bubble geometries to consider a priori because for each vacuum --- dS or AdS --- there are two basic objects. For dS these are the cap and bulb. For AdS, they are a finite patch containing the origin, and its infinite complement\footnote{By complement I mean the infinite portion of  AdS given by taking $\xi\in[\xi^*,\infty)$.}. We can make a candidate bounce geometry by connecting either of the AdS components to either of the dS components. The possibilities are shown in figure \ref{fig-mixed-cases}.

A configuration involving a patch connected to a cap, or a patch connected to a bulb, has finite action. One involving a complement connected to a cap, or a complement connected to a bulb, has infinite action. This means the former two combinations require the dS vacuum to be the parent, whereas the latter two require the AdS vacuum to be the parent (otherwise the resulting $B$ coefficients would not be finite). The two potential downtunneling transitions from dS are described by,
\begin{align}
B_{\text{patch-to-cap}}&=F_1(V_{\text{AdS}},x)+F_1(V_{\text{dS}},x)+B_{\text{wall}}-S_{\text{E}}[\text{dS}]\label{patch-to-cap-B}\\
B_{\text{patch-to-bulb}}&=F_1(V_{\text{AdS}},x)-F_1(V_{\text{dS}},x)+B_{\text{wall}}\label{patch-to-bulb-B}. 
\end{align}
All we've done here is make use of the fact that the Euclidean action from a finite patch of  AdS is still described by the $F_1$ function we defined in subsection \ref{subsec-intro-grav}. 

The two as of yet not eliminated uptunneling transitions are described by,
\begin{align}
B_{\text{cap-to-comp}}&=F_1(V_{\text{dS}},x)-F_1(V_{\text{AdS}},x)+B_{\text{wall}}\label{inf-to-cap-B}\\
B_{\text{bulb-to-comp}}&=F_2(V_{\text{dS}},x)-F_1(V_{\text{AdS}},x)+B_{\text{wall}}\\
&=S_{\text{E}}[\text{dS}]-(F_1(V_{\text{dS}},x)+F_1(V_{\text{AdS}},x))+B_{\text{wall}}\label{inf-to-bulb-B}. 
\end{align}
The analysis of these four conceivable transitions --- two downtunneling and two uptunneling --- is remarkably similar to the cases we've already investigated because the usual suspects appear: a sum or difference of cap functions, and the cubic wall term. To determine which of these has the capacity to produce an extremum in $B$, and under what conditions, we need only carry out the analysis we did in the dS-dS section, but for expressions involving mixed sign vacuum energies. 

Recall that cap functions are monotonically decreasing for all energies. Therefore, \eqref{inf-to-bulb-B} is the sum of three monotonically increasing functions and can have no nontrivial extremum. We rule out bulb-to-complement instantons just as we ruled out bulb-to-bulb instantons for dS-dS. 

Next, recall that the derivative  of a difference in cap functions is proportional to $x$ times $\Delta f$ evaluated with the energies in reverse order, as stated in equation \eqref{diff-cap-functions-deriv}, copied below for convenience,
\be
\frac{\partial (F_1(V,x)-F_1(\tilde{V},x))}{\partial x}=c_1 \kappa x(f(\tilde{V},x)-f(V,x)).
\ee
The same property we found for like-sign energies holds for mixed-sign energies: $f(\tilde{V},x)-f(V,x)<0$ and decreasing for $V<\tilde{V}$, and $f(\tilde{V},x)-f(V,x)>0$ and increasing for $V>\tilde{V}$. The effect is merely more pronounced for opposite sign energies. For example, consider negative $\tilde{V}=V_{\text{AdS}}$ and positive $V=V_{\text{dS}}$. This is the case relevant to cap-to-complement, \eqref{inf-to-cap-B}.  $f(V_{\text{AdS}},x)$ is monotonically increasing from $1$, and $-f(V_{\text{dS}},x)$ is increasing from $-1$. Therefore, $f(V_{\text{AdS}},x)-f(V_{\text{dS}},x)$ is monotonically increasing from zero, which means the derivative of $F_1(V_{\text{dS}},x)-F_1(V_{\text{AdS}},x)$ is positive at all nontrivial $x$ in the physical interval, and so $F_1(V_{\text{dS}},x)-F_1(V_{\text{AdS}},x)$ is monotonically increasing. Hence, $B_{\text{cap-to-comp}}$ in \eqref{inf-to-cap-B} is monotonically increasing, and cannot have a nontrivial extremum. Uptunneling from AdS is now completely ruled out.  

For the downtunneling transitions we have,
\be
B'_{\text{patch-to-cap}}=c_1 \kappa x (-2\bar{f}+\sigma \kappa x/2)\label{B'-patch-cap}
\ee
\be
B'_{\text{patch-to-bulb}}=c_1 \kappa x ( \Delta f+\sigma \kappa x/2)\label{B'-patch-bulb}
\ee
where $\Delta f=f(V_{\text{dS}},x)-f(V_{\text{AdS}},x)$, $2\bar{f}=f(V_{\text{dS}},x)+f(V_{\text{AdS}},x)$, and $x\in(0,\ell_{\text{dS}}]$. First note that the derivative of $Z=\frac{ \Delta f}{ \kappa \sigma x/ 2}$ is again negative everywhere in the interval, 
\be
\frac{\partial Z}{\partial x}= \frac{2}{\kappa \sigma x^2}\left( \frac{1}{\sqrt{1+\frac{x^2}{\ell^2_{\text{AdS}}}}}- \frac{1}{\sqrt{1-\frac{x^2}{\ell^2_{\text{dS}}}}}\right) <0
\ee
so the minimum value of $Z$ in the interval occurs at $x=\ell_{\text{dS}}$. The condition for the existence of a radius that extremizes $B_{\text{patch-to-bulb}}$  therefore is $Z(\ell_{\text{dS}})<-1$. Hence,
\be
f(V_{\text{dS}},\ell_{\text{dS}})-f(V_{\text{AdS}},\ell_{\text{dS}})<-\sigma \kappa \ell_{\text{dS}}/2
\ee
\be
\sigma<\frac{2}{\kappa}\frac{f(V_{\text{AdS}},\ell_{\text{dS}})}{ \ell_{\text{dS}}}= \frac{2}{\kappa} \sqrt{\frac{1}{\ell^2_{\text{dS}}}+\frac{1}{\ell^2_{\text{AdS}}}}.
\ee

For patch-to-cap, note that $-2\bar{f}$ once again starts off from $-2$, but is no longer necessarily either increasing or decreasing because $-f$ is monotonically increasing for the dS vacuum, but decreasing for the AdS. We should therefore group $-f(V_{\text{dS}},x)$ with the other increasing term, $\kappa\sigma x/2$, in the parentheses in \eqref{B'-patch-cap}. An extremum exists if,
\begin{align}
f(V_{\text{AdS}},x)&<\sigma \kappa x/2-f(V_{\text{dS}},x)\label{patch-cap-ineq}.
\end{align}

We can study two limiting cases, $\ell_{\text{dS}}\ll \ell_{\text{AdS}}$, and vice-versa. In the former case, we can use the small $x$ expansion for $f(V_{\text{AdS}},x)$, and find
\be
1+\frac{1}{2}\frac{x^2}{\ell^2_{\text{AdS}}}+\mathcal{O}((x/\ell_{\text{AdS}})^3)<\sigma \kappa x/2-f(V_{\text{dS}},x)
\ee
Evaluating at $x=\ell_{\text{dS}}$ we find,
\begin{align}
1+\frac{1}{2}\frac{\ell^2_{\text{dS}}}{\ell^2_{\text{AdS}}}+\mathcal{O}((x/\ell_{\text{AdS}})^3)<\sigma \kappa \ell_{\text{dS}}/2\\
\end{align}
hence, at leading order the condition,
\be
\sigma>\frac{2}{\kappa\ell_{\text{dS}}}\left(1+ \left(\frac{\ell_{\text{dS}}}{\ell_{\text{AdS}}}\right)^2\right).
\ee

In the $\ell_{\text{dS}}\gg \ell_{\text{AdS}}$ regime, we can instead use the asymptotic form of $f(V_{\text{AdS}},x)\sim x/\ell_{\text{AdS}}$. This gives, 
\be
0<x \left(\frac{\sigma \kappa}{2}-\frac{1}{\ell_{\text{AdS}}}\right)-f(V_{\text{dS}},x).
\ee
A simple way to guarantee this is satisfied is to impose $\sigma> \frac{2}{\kappa \ell_{\text{AdS}}}$. Notice the pleasant agreement in the form of the condition for the patch-to-cap extremum to exist in the two regimes,
\begin{align}
\sigma&>\frac{2}{\kappa\ell_{\text{dS}}},\quad \ell_{\text{dS}}\ll \ell_{\text{AdS}}\\
\sigma&>\frac{2}{\kappa\ell_{\text{AdS}}},\quad \ell_{\text{AdS}}\ll \ell_{\text{dS}}.
\end{align}

\section*{Acknowledgments}
I thank G. B. De Luca, M. Kleban, and A. Tomasiello for valuable discussions. My work is supported by the INFN.

\begin{appendix}
\section{Instanton Methods in Quantum Mechanics}\label{appendix-qm} 
\subsection{The decay rate}

\begin{figure}[b]
 \centering
\includegraphics[width=.5\linewidth]{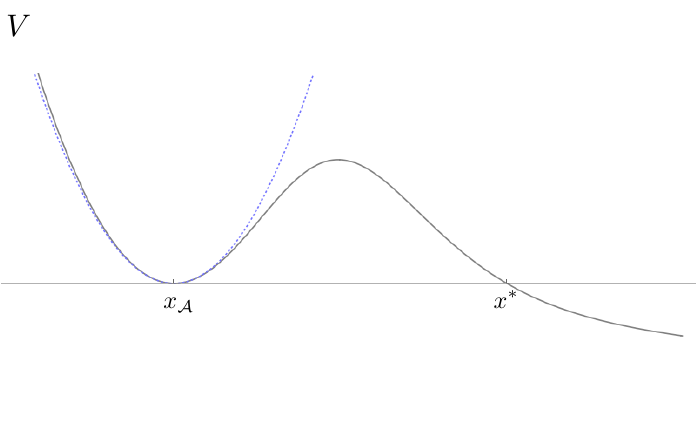}
\caption{\small The potential of a quantum mechanics system with a metastable state.}
 \label{fig-qm-V1-app}
\end{figure}

Consider a system of a quantum particle with potential $V(x)$ that has a local minimum at the  $x_{\mathcal{A}}$, like the  one in figure \ref{fig-qm-V1-app}. 
Roughly speaking, when the particle's wavefunction can  simultaneously be reasonably well-localized both in position and in momentum space about the classical equilibrium values $(x,p)=(x_{\mathcal{A}},0)$, then $$\langle V'(x)\rangle\approx V'({\langle x\rangle})=0. $$ The Ehrenfest theorem then implies that the expectation values $\langle x\rangle$ and $\langle p\rangle$ remain approximately constant, for at least some amount of time. 

\begin{figure}[h]
 \centering
\includegraphics[width=1\linewidth]{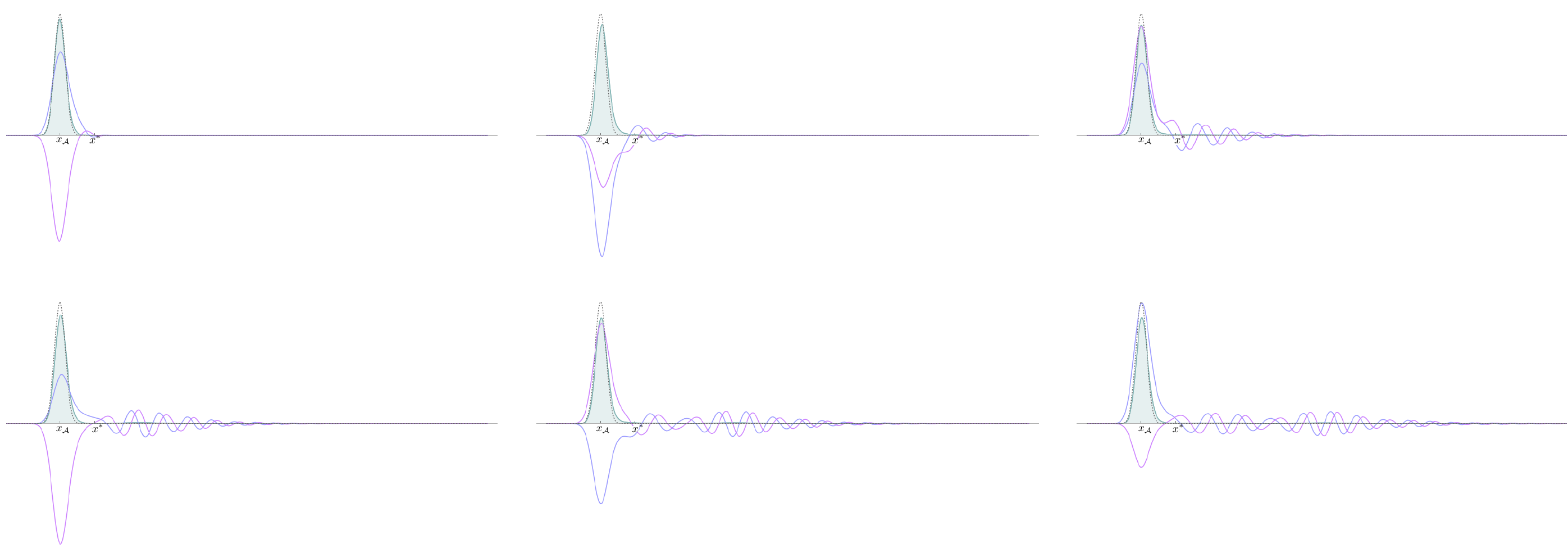}
\caption{\small Evolution of a Gaussian wavefunction initially centered at $x_{\mathcal{A}}$, specifically the ground state of the related harmonic oscillator potential. The initial probability density is shown by the black dotted curve, the time-dependent probability density is the green curve whose area is shaded, and the real and imaginary parts of the time-dependent wavefunction $\psi$ are shown in blue and pink.}
 \label{fig-qm-waves}
\end{figure}

The time-scale on which the state begins to deviate from the classical description is set by the rate at which the probability density integrated over the region where it initially had support,
$$P_{\mathcal{A}}=\int_{\mathcal{A}} dx |\psi(t,x)|^2$$
changes with time, where $\mathcal{A}$ denotes the neighborhood of $x_{\mathcal{A}}$. However, since an initially Gaussian probability amplitude will also oscillate within the well at frequency $\omega=\sqrt{V''(x_{\mathcal{A}})/m}$, we are really only interested in the dynamics of $P_{\mathcal{A}}$ smoothed over the time-scale $\Delta t=\omega^{-1}$; we are only interested in fluctuations on whom the state's fate depends. So, we define the decay rate,
\be
\Gamma(t)\equiv-\frac{1}{\tilde{P}_{\mathcal{A}}(t)}\frac{d\tilde{P}_{\mathcal{A}}(t)}{dt}\label{eq-gamma-def}
\ee
where  tilde indicates a quantity that is time-averaged over increments  $\Delta t$. The factor of inverse probability is included so that an exponentially decaying probability would result in a constant $\Gamma$, as this behavior is typical for decaying states. 
This definition can also be formulated in terms of the net flux $J(t,x)=\psi^\ast(t,x)\partial_x\psi-\psi(t,x)\partial_x\psi^\ast$ leaving the region $\mathcal{A}$ due to conservation of probability density 
\be
\Gamma(t)=\frac{1}{\tilde{P}_{\mathcal{A}}(t)}\int_{\partial \mathcal{A}} dx\thinspace \tilde{J}(x,t).
\ee

In the semiclassical approximation, instantons appear to provide a good estimate of the rate \eqref{eq-gamma-def} at the relevant times --- longer than $\omega^{-1}$, but shorter than the time-scale on which non-decay phenomena take place, for example reflection of probability amplitude back into the well. Instantons are particular, non-constant solutions to the Euclidean equations of motion that start and end in the false vacuum. The result of a path integral calculation carried out in Euclidean signature is a decay rate of the form,
\be
\Gamma= Ae^{-B/\hbar}\left[1+\mathcal{O}(\hbar) \right],\label{eq-decay-rate}
\ee
where $A$ and $B$ are constants fixed by the instanton associated with the decay channel, and the false vacuum. 

\subsection{Relation to the Euclidean path integral}
The derivation of \eqref{eq-decay-rate} begins from the Euclidean propagator, $K_{\text{E}}(x_0,x_1,T)$. This  is the extension of the ordinary propagator, $K(x_0,x_1,t)$, 
\be
K(x_0,x_1,t)=\langle x_1 | e^{-iHt/\hbar} |x_0\rangle\label{eq-lor-prop},
\ee
to values of $t$ on the negative imaginary axis. Here, $|x_0\rangle$ and  $|x_1\rangle$  are position eigenstates, and $H$ is the system's Hamiltonian operator. 

It makes sense to extend the time domain of the function \eqref{eq-lor-prop} because the $t\rightarrow -i \infty$ limit provides a means of extracting the ground state energy. To see this, expand $K(x_0,x_1,-iT)$ where $T$ is real and positive, in a set of energy eigenstates  $|\phi_n\rangle$,
\be
K_{\text{E}}(x_0,x_1,T)=\langle x_1 | e^{-HT/\hbar} |x_0\rangle
\ee
\be
K_{\text{E}}(x_0,x_1,T)=\sum_{n=0}^{\infty} e^{-E_nT/\hbar} \phi^\ast_n(x_1) \phi_n(x_0).
\ee
When $T$ is large this sum is dominated by the term with smallest $E_n$. It follows that,
\be
E_0= \lim_{T\rightarrow \infty} \left(-\frac{\hbar}{T}\log( K_{\text{E}}(x_0,x_1,T)  )\right).\label{eq-ground-state}
\ee

Now, the Euclidean propagator can also be expressed as a path integral,
\be
K_{\text{E}}(x_0,x_1,T)= N \int_{\text{Euc paths}} \mathcal{D}x \thickspace e^{-S_{\text{E}}[x]/\hbar}, \label{eq-E-path-int}
\ee
where $N$ is a normalization constant, the integral is over real functions on $[0,T]$ that are piece-wise $C^1$ that satisfy the boundary conditions $x(0)=x_0$, and $x(T)=x_1$, and $S_{\text{E}}$ is the (real) Euclidean action defined as,
\begin{align}
iS_{\text{Lor}}[x]&=\int_0^T d\tau \thickspace \frac{m}{2} \left(i \frac{dx}{d\tau}\right)^2-V(x)\\
&=-\int_0^T d\tau \thickspace  \frac{m}{2}  \left(\frac{dx}{d\tau}\right)^2+V(x)\\\label{euc-action-qm}
&\equiv - S_{\text{E}}[x].
\end{align}

The identification of a decay rate consists of extracting a particular set of contributions to $K_{\text{E}}(x_{\mathcal{A}},x_{\mathcal{A}},T)$ for large $T$ that arise within the context of the saddle approximation/method of steepest descent. Heuristically, 
\be
K_{\text{E}}(x_{\mathcal{A}},x_{\mathcal{A}},T)=N\sum_{n=0}^{n_{\text{max}}} I_n
\ee
where $n$ labels exact or approximate saddle points of $S_{\text{E}}$ satisfying the false vacuum boundary conditions, and $n_{\text{max}}$ may be infinite.

\subsection{On-shell paths}
The dominant term in $K_{\text{E}}(x_{\mathcal{A}},x_{\mathcal{A}},T)$ comes from the static false vacuum path, $x=x_{\mathcal{A}}$, because it has lowest $S_{\text{E}}$. Gaussian fluctuations about this path evaluate to,
\be
NI_0=Ne^{-S_{\text{E}}[x_{\mathcal{A}}]}\prod_{j}\int_{-\infty}^{\infty}(2\pi\hbar)^{-1/2}dc_j \thinspace e^{-\frac{1}{2}\lambda^{\text{f.v.}}_j c_j^2}=Ne^{-V_{\mathcal{A}} T/\hbar}\sqrt{\frac{1}{\prod_j\lambda^{\text{f.v.}}_j}}
\ee
where the $\lambda^{\text{f.v.}}_j$ are the eigenvalues of the operator appearing in the second variation of the Euclidean action, the term in parenthesis in,
\be
S''_{\text{E, f.v.}}(a,b)=\int_{0}^{T}d\tau \thinspace a(\tau) \left (-m\frac{d^2}{d\tau^2}+V''[x_{\mathcal{A}}]\right) b(\tau).\label{eq-sturm-liou-fv}
\ee
The eigenvalues, $\lambda^{\text{f.v.}}_j$, of the differential operator in parentheses in \eqref{eq-sturm-liou-fv} are all positive because $V''(x_{\mathcal{A}})/m=\omega^2>0$.
In the limit of large $T$, the inverse determinant factor in $NI_0$ behaves like,
\be
\lim_{T\rightarrow \infty}N \prod_j(\lambda_j^{\text{f.v.}})^{-1/2}=\left(\frac{\omega}{\pi\hbar}\right)^{1/2}e^{-\omega T/(2\hbar)}.
\ee
Hence, the contribution to $K_{\text{E}}(x_{\mathcal{A}},x_{\mathcal{A}},T)$ from Gaussian fluctuations about the static false vacuum path is,
\be
NI_0=\left(\frac{\omega}{\pi\hbar}\right)^{1/2}e^{-(V_{\mathcal{A}}+\frac{\hbar \omega}{2})T/\hbar}, \label{eq-zero-pt}
\ee
which results in the zero point correction to the ground state energy when used in \eqref{eq-ground-state}. 

The contribution to $K_{\text{E}}(x_{\mathcal{A}},x_{\mathcal{A}},T)$ relevant to the decay rate, on the other hand, comes from a portion of  function space surrounding the bounce,  $x_b(\tau)$. The bounce is a non-constant solution to the Euclidean equations of motion, 
\be
m\frac{d^2x}{d\tau^2}=\frac{d V}{dx},\label{eq-qm-Euc-EOM}
\ee
that satisfies the false vacuum boundary conditions, 
\be
x(0)=x(T)=x_{\mathcal{A}}.\label{eq-bcs-qm}
\ee
In addition to the trivial path that just sits at $x_{\mathcal{A}}$ (i.e. on top of the maximum of $-V$), there is a nontrivial solution to the boundary value problem, \eqref{eq-qm-Euc-EOM}-\eqref{eq-bcs-qm}, that explores only the barrier region. It is the path that starts at $x_{\mathcal{A}}$ with precisely the initial velocity $\dot{x}_0$ required to roll off the maximum (rightward), through the valley of $-V$ to a height $-V_{\mathcal{A}}+m\dot{x}_0^2/2$ in time $T/2$. The velocity at this point during the motion is momentarily zero because $m\dot{x}_0^2/2-V$  is conserved by \eqref{eq-qm-Euc-EOM}. The particle, whose position had been changing monotonically, reverses direction and rolls back towards $x_{\mathcal{A}}$ in a symmetric fashion, arriving (and technically passing) $x_{\mathcal{A}}$  exactly at time $T$. 

For large $T$, the requisite initial speed is small, and the particle bounces off the barrier nearly at the point of equipotential with $x_{\mathcal{A}}$, which we label $x^\ast$. At a certain point, increasing $T$ only serves to lengthen the flat initial and final segments of the trajectory, as the particle simply spends more time crawling off and up the local maximum of $-V$; the effect on the nontrivial part of the trajectory (the bump) is  insignificant. As mentioned in section \ref{intro}, as $T\rightarrow\infty$ the difference in Euclidean action between the bounce and  $S_{\text{E, f.v.}}$ approaches 
\be
 B=  \int_{x_{\mathcal{A}}}^{x^\ast} dx\sqrt{2m(V(x))-V_{\mathcal{A}})}.\label{eq-B-qm}
\ee 

\begin{figure}
 \centering
\includegraphics[width=1\linewidth]{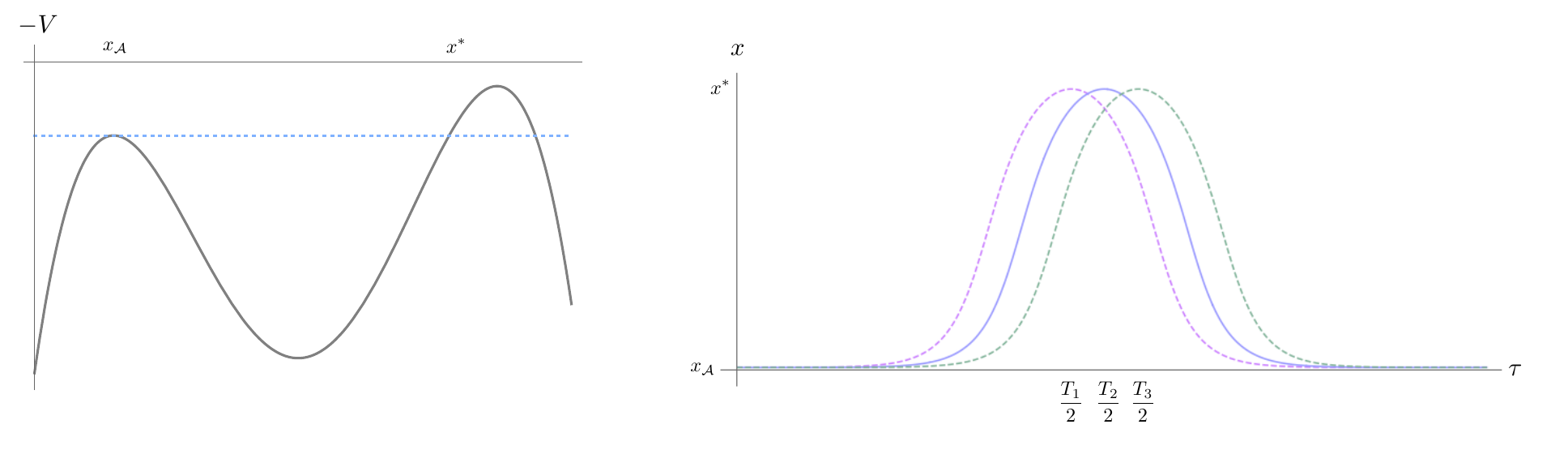}
\caption{\small Left: inverted potential. Right: on-shell trajectories starting at $x_{\mathcal{A}}$ with small, but distinct, initial velocities. The  nontrivial portions of the trajectories --- the bumps --- are the same. Decreasing/increasing the initial rightward velocity just displaces the center of the bounce $T/2$.} \label{fig-qm-bounces}
\end{figure}

Since the width $\Delta \tau$ of the bounce is finite, multi-bounce paths can be constructed by adding together $n<\frac{T}{\Delta \tau}$ single bounces whose centers have been shifted to any desired locations in $[0,T]$ so long as no two bumps overlap, and then shifting the sum vertically by $(n-1)x_{\mathcal{A}}$ to satisfy the false vacuum boundary conditions. In other words add $n$ horizontally shifted bump profiles, where a bump is defined as $x_b(\tau)-x_{\mathcal{A}}$, and add to this superposition the constant $x_{\mathcal{A}}$. This is an $n$-bounce; it is an approximate fixed point of the Euclidean action and has Euclidean action $nB+S_{\text{E}}[x_{\mathcal{A}}]$.

The eigenfunctions $\{y_n\}_{n=0}^{\infty}$ of the differential operator inside $S_{\text{E}}''[x_b]$,
\be
-m\frac{d^2}{d\tau^2}+V''(x_b(\tau)),\label{schro-op}
\ee
form a basis for the function space $L^2$ because the operator is Sturm-Liouville. 
 \eqref{schro-op} is a Schr\"{o}dinger-type operator in one dimension, with $\tau\in [0,T]$ playing the role of the  spatial variable, and $V''(x_b(\tau)) $ playing the role of the potential (up to factors of $m$). The relevant boundary conditions for the $y_n$ are Dirichlet, because  we'll use these eigenfunctions to expand $S_{\text{E}}[x]$ about $x_b$, which already satisfies the boundary conditions \eqref{eq-bcs-qm}. Outside of the bump portion of the bounce, $V''(x_b(\tau))$ is exponentially close to $m\omega^2>0$. At the center of the bounce $V''(x_b(\tau))=V''(x^\ast)>0$. In between, however,  $V''(x_b(\tau))$ passes through  the value of $V''$ at top of the barrier, which is negative, $V''_{\text{peak}}<0$. Because the full bounce is time-reversal invariant, and  each half of $x_b$ is smooth and monotonic, $V''(x_b(\tau))$ takes the form of a symmetric finite double-well Schr\"{o}dinger potential with negative minimum. 
 
Coleman's clever reasoning goes as follows. The function $\frac{dx_b}{d\tau}$ is an eigenfunction of \eqref{schro-op} with eigenvalue zero because of the Euclidean equations of motion,
\begin{align}
\left[-m\frac{d^2}{d\tau^2}+V''(x_b(\tau))\right]\frac{dx_b}{d\tau}&= \frac{d}{d\tau}\left[-m\frac{d^2 x_b}{d\tau^2}+V'(x_b)\right]=0
\end{align}
$\frac{dx_b}{d\tau}$ is odd and vanishes at the center of the bounce. This is its only node, implying there is precisely one eigenfunction  of \eqref{schro-op} with lower --- hence negative --- eigenvalue. 

We arrange the eigenvalues in ascending order starting from that of the nodeless negative mode, $\lambda_0<0$. $\lambda_1=0$, and the normalized zero mode, $y_1=B^{-1/2}\frac{dx_b}{d\tau}$ (a consequence of the equations of motion). Coleman's idea is to include the Gaussian fluctuations about each multi-bounce by using $n$ copies of single bounce modes, as this set well approximates the actual eigenmodes of the operator in $S_{\text{E}}[x_{n\text{-bounce}}]$ when the $n$ bounces are well-separated. Ultimately, $T$ will be taken to $\infty$ as a calculational technique, which implies there will be arbitrarily many $n$-bounces with arbitrarily well-separated separated bounces.

\begin{figure}
 \centering
\includegraphics[width=1\linewidth]{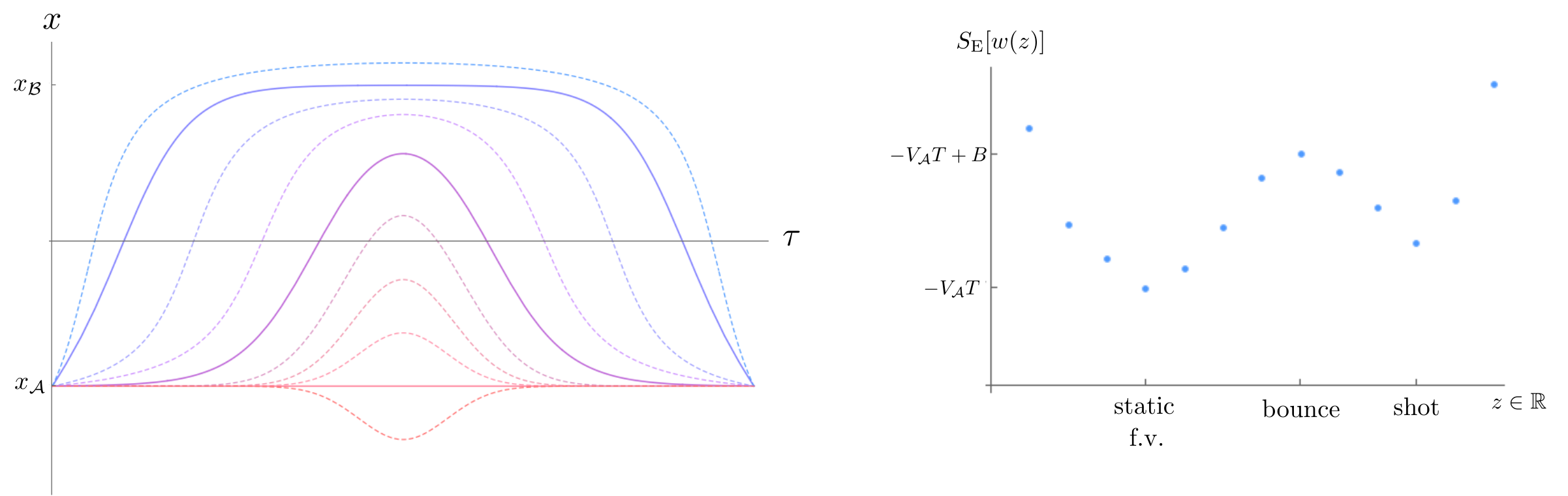}
\caption{\small Left: Set of paths with $x(0)=x(T)=x_{\mathcal{A}}$.  For a double-well potential with two distinct vacua, on-shell paths are shown by solid lines: red is the static false vacuum path, purple the bounce, and blue the "shot."  The deformation of the static false vacuum path to the shot corresponds to a moving along a curve $w(z)$ in function space that passes through all three saddles, and specifically through the bounce in the direction of its negative mode. The negative mode direction corresponds to contracting/expanding the bounce. The negative mode can be seen in the adjacent plot. Right: Euclidean actions for paths along $w(z)$. The shot is a local \emph{maximum} of $S_{\text{E}}$.}
 \label{fig-paths}
\end{figure}

If $V$ has no local minimum on the other side of the barrier, like in figure \ref{fig-qm-V1-app}, the bounce is the only on-shell path in addition to the static false vacuum path satisfying the boundary conditions \eqref{eq-bcs-qm}. If instead there is a lower $V$ local minimum on the other side of the barrier, for example as is shown on the left in figure \ref{fig-qm-bounces}, then there is another nontrivial on-shell path in addition to the bounce, appropriately dubbed ``the shot" by \cite{Andreassen:2016cvx}. This is a path that starts at $x_{\mathcal{A}}$ with much larger kinetic energy that the bounce. It's initial rightward speed $\dot{x}_0$ is tuned to be just less that $\sqrt{2m(V_{\mathcal{A}}-V_{\mathcal{B}})}$. Consequently, the particle overshoots $x^\ast$, and comes nearly to rest at the top of the hill of the inverted potential at $x_{\mathcal{B}}$, the higher of the two hills. For any value of $T$, there is one value for the initial kinetic energy $\frac{m}{2}\dot{x}_0^2$ close to $V_{\mathcal{A}}-V_{\mathcal{B}}$ that has the particle roll back to  $x_{\mathcal{A}}$ at time $T$. The three on-shell paths for the double-well potential in figure \ref{fig-qm-bounces} are shown in the plot on the left-hand side of figure \ref{fig-paths}. The shot is the solid blue curve, the bounce is solid purple, and the static false vacuum is solid red.  A thorough presentation of the steepest descent procedure, and its application in the study of vacuum decay can be found in \cite{Andreassen:2016cvx}. For the purposes of providing a self-contained discussion here, I include figure  \ref{fig-paths}  as a similar version of the plots in their figure 6. Finally, going forward we take the particle to have unit mass, $m=1$. It can be restored at the end by dimensional analysis.

\subsection{Saddle Approximation}
Extending Laplace's method/steepest descent to infinite dimensional integrals presents ambiguities in general. Steepest descent is a technique for approximating 1-dimensional integrals, for example of the form,
\be
I(M)=\int_{-\infty}^{\infty}e^{M f(x)} g(x)dx
\ee
where $M$ is a large constant. The idea is to exploit the freedom granted by Cauchy's theorem to deform the contour of integration into the complex plane within a region where integrand is an analytic function. Analyticity results in several important factors which together make a particular choice of the contour, that of ``steepest descent", a systematic means of estimating the integral by producing an asymptotic series in $\frac{1}{M}$. 

For complex arguments $z=x+i y$, the functions $f$ and $g$ are in general complex. The gradients of $f_{\text{R}}(x,y)\equiv\text{Re}(f(x+iy))$ and $f_{\text{I}}(x,y)\equiv\text{Im}(f(x+iy))$ in the $x$-$y$ plane are two real vector fields. The Cauchy-Riemann equations imply they are orthogonal to each other at points $(x,y)$ where $f(z)$ is analytic. The strategy is to first restrict the choice for the contour $C$ to those along which $ f_{\text{I}}$ remains constant. This eliminates the complication presented by the otherwise oscillatory factor $e^{ i M f_{\text{I}}(x,y)}$. Parameterizing the position along such contour as $z(t)=x(t)+iy(t)$ gives,
\begin{align}
I(M)&=\int_{C}^{}e^{M f(z)} g(z)dz\\
&= e^{iM f_{\text{I}}}\left( \int_{a}^{b}e^{M f_{\text{R}}(x(t),y(t))}  \tilde{g}_{\text{R}}(t)dt + i\int_{a}^{b}e^{M f_{\text{R}}(x(t),y(t))}   \tilde{g}_{\text{I}}(t) dt \right)\label{eq-laplace-integrals}
\end{align}
where $\tilde{g}_{\text{R}}(t)=z'(t)\text{Re}(g(z(t)))$, $\tilde{g}_{\text{I}}(t)=z'(t)\text{Im}(g(z(t)))$. Since $f_{\text{I}}$ is constant along $C$, the gradient of $ f_{\text{R}}$ always points either in the same direction of the tangent vector $x'(t)  \hat{\mathbf{i}}+y'(t)\hat{\bold{j}}$, or exactly opposite to it, i.e. in the direction $-x'(t) \hat{\mathbf{i}}-y'(t)\hat{\mathbf{j}}$.  We'll only ever be interested in $g(z)=1$ so we eliminate it now to simplify the notation.

Another consequence of the analyticity of $f(z)$ is that the only extrema possible of $f_{\text{R}}(x,y)$ are saddle points (for example, if $\partial^2_xf_{\text{R}}>0$ at a zero of $\nabla f_{\text{R}}$,  then $\partial^2_yf_{\text{R}}$ must be negative because $f_{\text{R}}$ is harmonic). Note that if $\nabla f_{\text{R}}$ vanishes at a point $(x_0,y_0)$, so does $\nabla f_{\text{I}}$ by the Cauchy-Riemann equations. Hence, such a point corresponds to a critical point $z_0=x_0+iy_0$ of $f(z)$ because,
\be
\frac{df}{dz}=(\partial_x -i\partial_y)(f_{\text{R}}(x,y)+if_{\text{I}}(x,y)).
\ee

The prescription is to select the constant $f_{\text{I}}$ contour that permits each of the integrals in \eqref{eq-laplace-integrals} to be estimated using Laplace's method. This means choosing a contour $C$ that passes through at least one saddle point of $f(z)$, and does so in the direction of maximal decrease of $|f_{\text{R}}|$ away from the saddle. A portion of the original contour is deformed (if necessary) such that tangents to the curve are aligned with $\nabla f_{\text{R}} $ as a saddle point is approached, and exactly opposed as the saddle is departed. This procedure is applied until the entire contour reaches the original endpoints. The result is a contour $C$ that everywhere lies along the \emph{steepest descent} emanating from any and all saddles it encounters. Note that the criteria of steepest descent allows for $C$ to cross more than one saddle point of $f(z)$, yet does not require it cross all.

In the neighborhood of an extremum $(x_0, y_0)$ with non-vanishing second derivatives, the local steepest descent deformation is accomplished as follows.
Expand $f$ in a Taylor series about a saddle $z_0$, with the displacement from $z_0$ expressed as as $re^{i\theta}$, and let $\alpha$ denote the phase of $\frac{d^2f}{dz^2}|_{z_0}$,
\begin{align}
f(z)&=f(z_0)+\frac{1}{2}|f''(z_0)|e^{i\alpha}r^2e^{2i\theta}+\mathcal{O}((z-z_0)^3)\\
\text{Im}(f(z))&=\text{Im}(f(z_0))+ \frac{1}{2}|f''(z_0)| r^2 (\cos(2\theta+\alpha)+i\sin(2\theta+\alpha))\\
\text{Im}(f(z))&=\text{const}\\
\rightarrow 2\theta+\alpha &= k\pi,\thickspace k\in \mathbb{Z}
\end{align}
The descent directions passing through the saddle are those rays at angle, 
$$\theta=\frac{1}{2}(k \pi -\alpha)$$ with respect to the $x$-axis, where 

whose $k$ yields an integrable expression in the leading order contribution to $I(M)$ from the saddle region, 
\be
\int_{C_\text{saddle}}e^{M f(z)}dz \approx e^{Mf(z_0)}\int_0^{\infty} e^{i\theta} dr e^{\frac{M}{2} |f''(z_0)| \cos(k\pi) r^2}. 
\ee
In other words, odd $k$,
\be\label{eq-half-gaussian-result}
\int_{C_\text{saddle}}e^{M f(z)} dz\approx e^{Mf(z_0)} e^{i ( \pi-\alpha)/2} \frac{1}{2} \sqrt{\frac{2\pi}{ M |f''(z_0)|}} 
\ee
Hence, for every quadratic critical point of $f$ there are two distinct descent directions, $\theta^{\text{desc}}_\pm=\pm \pi/2 - \alpha/2$. Locally, the  integrand  looks identical in these directions. Similarly, there are two ascent directions at $\theta^{\text{asc}}_\pm=\pm \pi - \alpha/2$. Equation \eqref{eq-half-gaussian-result}  shows that for any kind of saddle with non-trivial $f''$, the leading order contribution to $I(M)$ coming from following one of the saddle's descent directions is the usual half-Gaussian result, times an overall phase, $e^{i ( \pi-\alpha)/2}$. 

The kinds of saddles that are relevant to us are those which lie on the real-line, and are critical points of a real function $f(z)=-S_{\text{E}}$. There are two types of real critical points, $z_0=x_0$. Either $f''(x_0)<0$ so the saddle is a maximum of $S_{\text{E}}$ in the real direction, and minimum in the imaginary direction, or $f''(x_0)>0$ and the saddle is a minimum in the real direction and maximum in the imaginary direction. The first type corresponds to $\alpha \pmod{2\pi}=\pi$ and requires no deformation away from the real-axis. Indeed, we see the phase in its \eqref{eq-half-gaussian-result} is zero, and we thus recover the correct result real. The second kind of saddle, has $\alpha \pmod{2\pi}=0$, and produces an extra factor of $e^{\pm i \pi/2}=\pm i$, as compared to the real result. (if $g(z)\neq 1$ the result \eqref{eq-half-gaussian-result} simply comes with a factor of $g(z_0)$.)

It is important to underscore that \eqref{eq-half-gaussian-result} only reflects the leading order contribution from the portion of $C$ passing through $z_0$, and in one direction. To recover the asymptotic series, contributions from all portions of the descent contour need to be included, and computed consistently to a given order in $1/M$. 
The contribution from the vicinity of the saddle can be computed to any order in $\frac{1}{M}$ by using a Taylor expansion for $f$ to a given order to solve for the higher order corrections to the constant $f_{\text{I}}$ curves emanating from the saddle. Which curves correspond to descent directions is determined by plugging the Taylor expansion for $f$  into the integrand. The contribution from each segment of $C$ lying along one of these higher-order corrected descent directions can be evaluated because the deformation  is precisely that which ensures the integrability of $e^{Mf(z)}$ with $f$ expanded to the relevant higher-order.

\subsection{Applying steepest descent to the path integral}
Coleman's calculation applies the technique of steepest descent to each negative mode direction emanating from each saddle independently, essentially by using an iterative procedure. Consider first a curve in function space, $w(z)$, which passes through the static false vacuum path, say at $z=0$, and then successively through each $n$-bounce saddle beginning with $n=1$ in the direction of its \emph{first} negative mode out of $n$, i.e. the negative mode associated with the first of its $n$ far-separated bounces, $y_0(\tau)$. 
The function $S_{\text{E}}[w(z)]$ has a local minimum at $z=0$, and a local maximum at each $z$ value indicating the location of an $n$-bounce. The first step is to apply steepest descent to the 1-dimensional integral over $z$,
\begin{align}
K_{\text{E}}(x_{\mathcal{A}},x_{\mathcal{A}},T)&=N'\int \mathcal{D}w(z)^\perp \int_{-\infty}^{\infty}  dz \thinspace e^{-S_{\text{E}}[w(z)+ w(z)^\perp]/\hbar}
\end{align}
We are free to choose the $z$-values where the $n$-bounces are located along the curve $w(z)$ because the overall scale of $z$ is irrelevant. It is accounted in the proportionality coefficient when expressing the measure $\mathcal{D}x\propto dz$. We choose the natural numbers, meaning the single bounce is located at $z=1$, the double-bounce at $z=2$, and so on.  

The Euclidean action increases in all directions away from the static false vacuum path, so no deformation of the $z$-integral is required around $z=0$. Hence, the leading order contribution from the false vacuum saddle in steepest descent just produces the Gaussian fluctuations we already computed (when orthogonal directions are integrated over). As we proceed along the negative real axis we approach the single-bounce, and $w$ takes the form $x_b(\tau)+c_0 y_0(\tau)$. $-S_{\text{E}}$ decreases up until we reach the bounce. If we attempt to compute its leading contribution, we encounter half of a wrong-sign Gaussian integral, as expected,
\begin{align}
&N' \int \mathcal{D}w(z)^\perp \int_{1}^{1+\Delta z}  dz  e^{-S_{\text{E}}[w(z)+ w(z)^\perp]/\hbar}\\
 &\approx N \int \mathcal{D}w(z)^\perp \int_{-\infty}^{\infty} dc_0 (2\pi \hbar)^{-1/2} e^{-S_{\text{E}}[x_b+c_0 y_0+ w(1)^\perp ]/\hbar}\\
&=N  e^{-S_{\text{E}}[x_b]/\hbar} \int \mathcal{D}w(1)^\perp e^{-S_{\text{E}}[w(1)^\perp]/\hbar}   \int_{-\infty}^{\infty} dc_0 (2\pi \hbar)^{-1/2} e^{+\frac{1}{2\hbar}|\lambda_0|c_0^2}.
\end{align}
We trade this contribution for that computed using one of the bounce's descent rays. Namely, the factor from orthogonal directions times $\pm N  e^{-S_{\text{E}}[x_b]/\hbar} J$, where $J$ is given by, 
\begin{align}
J\equiv & (2\pi \hbar)^{-1/2}\int_0^{+ i \infty} dc_0  \thinspace  e^{+\frac{1}{2\hbar}|\lambda_0| c_0^2}\\
&=(2\pi \hbar)^{-1/2} \int_0^{+ \infty} i d \eta \thinspace e^{-\frac{1}{2\hbar}|\lambda_0| \eta^2}\\
&=\frac{i}{2}\frac{1}{\sqrt{|\lambda_0|}}.
\end{align}
\begin{figure}
 \centering
\includegraphics[width=1\linewidth]{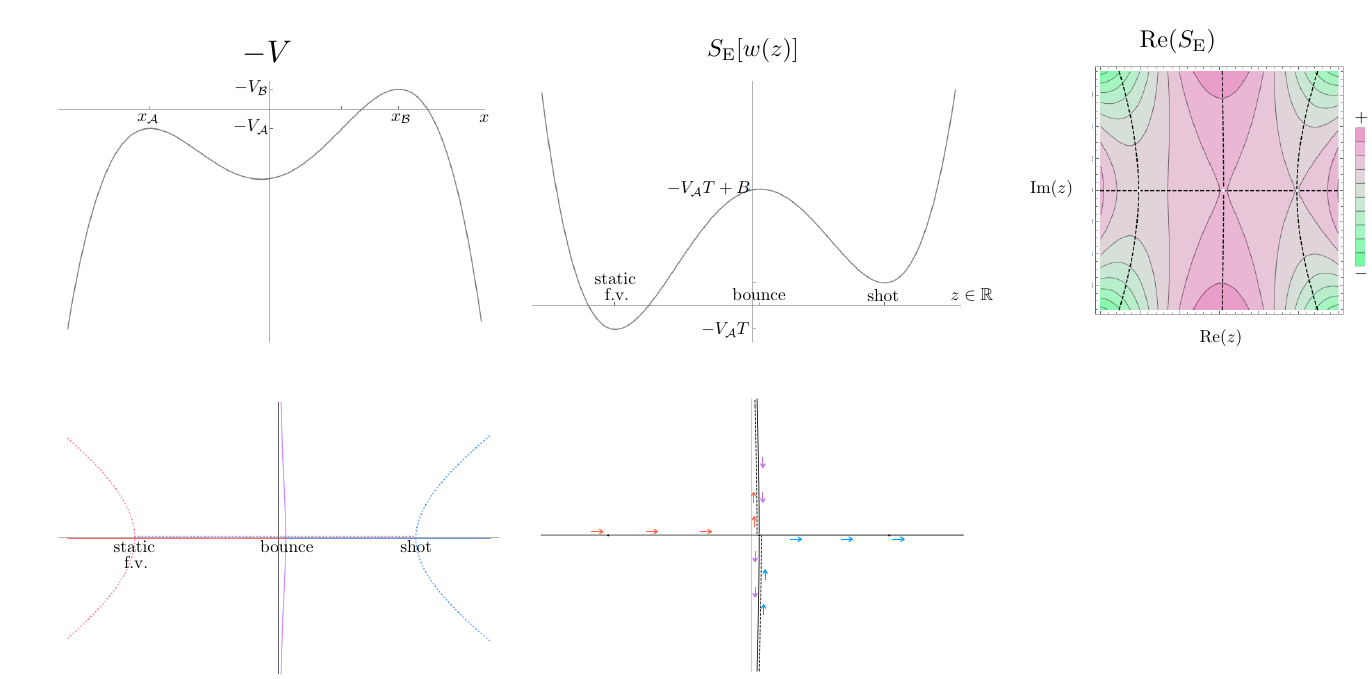}
\caption{\small \emph{Top row:} Left: Potential with two local minima. Middle: a real quartic function $h(x)=(x^2-1)^2+\frac{x}{5}$ that serves as a stand-in for $S_{\text{E}}[w(z)]$, plotted as a function of real $z$. Right: Contour plot of $\text{Re}(h(z))$ in the complex $z$-plane. \emph{Bottom row:} Left: curves in the complex plane where $\text{Im}(h(z))=0$. Solid lines represent descent contours emanating from a saddle, dotted lines represent ascent directions. The different saddles are distinguished using different colors. Along the x-axis where the ascent directions emanating from one saddle lead into/coincide with the descent directions of a neighboring saddle, the two curves are offset for legibility. Right: Full steepest descent contour for $\int_{-\infty}^{\infty}e^{-h(x)}$. Follow the red arrows first, then purple, then blue.}
 \label{fig-contours}
\end{figure}
Each portion of the real $z$-axis around the location an $n$-bounce encounters the same wrong-sign Gaussian,
\be
\int_{n}^{n+\Delta z}  dz  e^{-S_{\text{E}}[w(n)+ w(n)^\perp]/\hbar} \approx e^{-S_{\text{E}}[x_{n-\text{bounce}}]/\hbar }e^{-S_{\text{E}}[w(n)^\perp]/\hbar}   \int_{0}^{\infty} dc_0 (2\pi \hbar)^{-1/2} e^{+\frac{1}{2\hbar}|\lambda_0|c_0^2}.\nonumber
\ee
We therefore we make the same deformation of the contour into the complex $z$-plane as we did for the single-bounce. Our convention is to select the bounce's descent ray in the upper half-plane (but the choice is arbitrary). So far we have,
\begin{align}
\int_{-\infty}^{\infty}  dz &\rightarrow \left( \int_{-\infty}^{1}  dz + \int_{C_1} + \int_{C_2} +\dots + \int_{C_{n_{\text{max}}}} \right ) \nonumber\\ 
K_{\text{E}}(x_{\mathcal{A}},x_{\mathcal{A}},T)&=NI_0+\dots+ N e^{-S_{\text{E}}[x_b]/\hbar} \int \mathcal{D}w(1)^\perp e^{-S_{\text{E}}[w(1)^\perp]/\hbar} (J+\dots)\nonumber\\
&\qquad \qquad +  \sum_{n=2}^{n_{\text{max}}} N e^{-S_{\text{E}}[x_{n\text{-bounce}}]/\hbar} \int \mathcal{D}w(n)^\perp e^{-S_{\text{E}}[w(n)^\perp]/\hbar} (J+\dots) \nonumber
\end{align}
where the ellipses following $NI_0$ represent subleading perturbative terms in the static false vacuum contribution, and those following the $J$'s represent subleading corrections to $J$ as well as integrals over any remaining segments of the $C_i$. For example, if after the bounce $w(z)$ passes through a local minimum of $S_{\text{E}}$, for instance through the shot saddle point, the full deformation around the bounce involves two additional segments which precisely cancel $J$,
\be
\int_{C_1}dz \approx \int_1^{1+i\infty}dz + \int_{1+i\infty}^{1-i\infty}dz + \int_{1-i\infty}^{1}dz + \int_{C_\text{shot}} dz.
\ee 
This contour is shown in the figure \ref{fig-contours}. The first three segments produce the following leading order contribution,
\be
N e^{-S_{\text{E}}[x_b]/\hbar} \int \mathcal{D}w(1)^\perp e^{-S_{\text{E}}[w(1)^\perp]/\hbar} (J-2J+J) 
\ee
which vanishes. This is to be expected for a real integral $\int_{-\infty}^{\infty} e^{-S_{\text{E}}[w(z)]/\hbar}$, since ultimately all imaginary terms in the saddle approximation would need to cancel order by order. (the ground state energy is  real, after all.) Nevertheless, the decay rate is related specifically to the magnitude the contribution from \emph{one} of the bounce's descent rays.

Now we will evaluate the $\int \mathcal{D}w(1)^\perp$ factor in the single-bounce contribution. The orthogonal directions lie in the span of the eigenmodes of the operator in $S''_{\text{E}}[x_b]$ excluding the negative mode direction, $w(1)^\perp= \text{span}(\{y_n\}_{n=1} ^{\infty})$. Thus, we find the leading contribution from the single bounce coming from the imaginary half-line segment of steepest descent contour $C_1$ is, 
\begin{align}
&N e^{-S_E[x_b]} \frac{i}{2}\frac{1}{\sqrt{|\lambda_0|}} \int (2\pi \hbar)^{-1/2} dc_1 \prod_{j=2}^{\infty} \int_{-\infty}^\infty dc_j (2\pi \hbar)^{-1/2}e^{-\frac{1}{2\hbar}\lambda_j c_j^2} \\
&=N e^{-S_E[x_b]} J \thickspace (\text{Zero mode factor}) \times (\text{positive modes factor}) 
\end{align}
\be
(\text{Zero mode factor}) \times (\text{positive modes factor})=\frac{\text{Range}(c_1)}{ (2\pi \hbar)^{1/2}}\thinspace\prod_{n>1}\lambda_n^{-1/2}.
\ee
$\text{Range}(c_1)$ works out to $T\sqrt{B}$ due to the normalization of $y_1(\tau)$ we mentioned earlier. 

Now the idea is to deal with all the second negative mode directions for the $n\geq2$ multi-bounces. Replacing all the second wrong sign Gaussian integral in the remaining $n$-bounce contribution with $(J+\dots)$, and then all the third wrong sign Gaussians, and so on yields,
\begin{align}
K_{\text{E}}(x_{\mathcal{A}},x_{\mathcal{A}},T)&=NI_0+\dots+ N e^{-S_{\text{E}}[x_b]/\hbar}  \left[ \frac{i}{2}\left(T\sqrt{\frac{B}{2\pi \hbar}}\right) (\text{det}'(-d_\tau^2+V''[x_b]))^{-1/2}+\dots) \right]\nonumber\\
&+  \sum_{n=2}^{n_{\text{max}}} N e^{-S_{\text{E}}[x_{n\text{-bounce}}]/\hbar}  (J+\dots)^n  \prod_{j=1}^n \int (2\pi \hbar)^{-1/2} dc^{(j)}_1 \left(\prod_{k>1}\lambda_k^{-1/2}\right)^n\nonumber
\end{align}
where  $\text{det}'$ is the determinant factor obtained by omitting zero eigenvalues, and taking the absolute value of negative eigenvalues. This procedure is iterative in the sense that it is equivalent to applying steepest descent on the set of 1-dimensional real integrals arising from evaluating the functional $S_{\text{E}}$ along $n$ curves $w_{(j)}(z)$ in function space that pass through the false vacuum static path, and then through each multi-bounce saddle in the direction of the negative mode associated with its $j^{\text{th}}$ out of $n<n_{\text{max}}$ bounces. 

The volume factor from integrating over the zero modes for the $n$-bounce works out to
\be
 \prod_{j=1}^n \int (2\pi \hbar)^{-1/2} dc^{(j)}_1=  \frac{1}{n!}\left(T\sqrt{\frac{B}{2\pi \hbar}}\right)^n
\ee
The volume isn't simply the 1-dimensional result raised to the $n$ because the range of each $c^{(j)}_1$ is limited by the fact that it cannot correspond to shifting a bounce past one of its neighbors. The bounces are indistinguishable, so doing so would amount to double counting the contribution to the Euclidean propagator.
Putting everything together gives the decomposition,
\begin{align}
K_{\text{E}}(x_{\mathcal{A}},x_{\mathcal{A}},T)&=NI_0+\dots +N e^{-S_{\text{E}}[x_{\mathcal{A}}]/\hbar} \sum_{n=1}^{n_{\text{max}}}  \frac{e^{-nB/\hbar}}{n!}   \left[ \frac{iT}{2}\sqrt{\frac{B}{2\pi \hbar}} (\text{det}'(-d_\tau^2+V''[x_b]))^{-1/2}+\dots) \right]^n\nonumber\\
&= NI_0 \left( 1+\dots + \sum_{n=1}^{n_{\text{max}}}  \frac{1}{n!}   \left[  \frac{i}{2} e^{-B/\hbar}\left(T\sqrt{\frac{B}{2\pi \hbar}}\right) \left(\sqrt{\frac{\text{det}'(-d_\tau^2+\omega^2)}{\text{det}'(-d_\tau^2+V''[x_b])}}+\dots\right) \right]^n  \right).\nonumber
\end{align}
As $T$ goes to $\infty$, the number of bounces that can fit in the Euclidean time interval becomes infinite. The small imaginary contribution to $K_{\text{E}}(x_{\mathcal{A}},x_{\mathcal{A}},T)$ proportional to $J$ coming from the neighborhood of the single bounce as it is approached from the false vacuum therefore exponentiates when the analogous partial neighborhoods' contributions of multi-bounces are summed over,
\begin{align}
K_{\text{E}}(x_{\mathcal{A}},x_{\mathcal{A}},T)&=NI_0 e^{ \frac{i}{2} e^{-B/\hbar}\left(T\sqrt{\frac{B}{2\pi \hbar}}\right) \left(\sqrt{\frac{\text{det}'(-d_\tau^2+\omega^2)}{\text{det}'(-d_\tau^2+V''[x_b])}}\right)}+\dots\label{eq-exponentiate}
\end{align}
where the ellipses represent subleading contributions and the portions of the steepest descent contour excluded. 

Define the constant,
\be
A\equiv \sqrt{\frac{B}{2\pi \hbar}} \left(\sqrt{\frac{\text{det}'(-d_\tau^2+\omega^2)}{\text{det}'(-d_\tau^2+V''[x_b])}}\right).
\ee
In the sense of equation \eqref{eq-ground-state}, the net effect of the $\propto J^n$ contributions shown in \eqref{eq-exponentiate} can be viewed as an exponentially small imaginary correction of the form $-\frac{i}{2} A e^{-B/\hbar}$  to the ground state energy. It is exponentially suppressed relative to  the zero point correction $V_{\mathcal{A}}+\hbar \omega/2$. This imaginary ``energy" term is then interpreted in the spirit of the poles and analytic structure of the function, 
\be
\text{Tr}\left(\frac{1}{zI- H}\right).
\ee
where $H$ is the quantum operator acting on a Hilbert space. A pole that lies off the real axis is interpreted as representing an instability occurring with/described by the rate $\Gamma=-2\text{Im}(\tilde{E})$ where $\tilde{E}$ is the location of the pole in the complex $z$-plane. In this way, we arrive at the identification
\be
\Gamma=Ae^{-B/\hbar}.
\ee

The fact that $B$, the difference between the bounce and static false vacuum actions worked out to \eqref{eq-B-qm} means that this procedure reproduces the correct exponential factor expected from a WKB calculation of the transmission coefficient. As mentioned in the introduction, this provides strong evidence for trusting that instantons in fact calculate semi-classical decay rates. A fully rigorous explanation for why the particular contributions to $K_{\text{E}}(x_{\mathcal{A}},x_{\mathcal{A}},T)$ we've retained (the imaginary half Gaussians $\sim J$ from each negative mode of the $n$-bounces) should be linked to the quantity of interest in vacuum decay, \eqref{eq-gamma-def}, is (as far as I can see) unclear. 

In specific cases, like the symmetric double-well potential for the quantum particle, a more direct line can be traced, essentially by relating $\Gamma$ to a difference in energies \cite{Coleman:1978ae}. It may well be that some of the ambiguities regarding the interpretation of negative modes for gravitational instantons, particularly within the context of decay from dS parents, might be clarified by a more scrupulous application/understanding of steepest descent.

\end{appendix}

\bibliographystyle{klebphys2}
\bibliography{tunnelingrefs}
\end{document}